\documentclass[%
superscriptaddress,
 amsmath,amssymb,
 aps,
 reprint,
prb,
floatfix,
]{revtex4-2}
\usepackage{graphicx,xcolor}
\definecolor{darkblue}{RGB}{0,0,150}
\definecolor{nightblue}{RGB}{0,0,100}

\usepackage{amsmath, amssymb, physics}

\usepackage{mathrsfs,dsfont,mathtools}
\usepackage{graphicx}

\usepackage{dcolumn}
\usepackage{bm}
\usepackage[
colorlinks,
citecolor=darkblue,
linkcolor=darkblue,
urlcolor=nightblue, breaklinks=true]{hyperref}

\usepackage[english]{babel}
\usepackage[babel,kerning=true,spacing=true]{microtype}
\usepackage{feynmp-auto}

\AtBeginDocument{\renewcommand{\natexlab}[1]{#1}}
\begin{document}

\title{Tunable Skyrmions in a Topological Wigner Crystal}

\author{David A. Dahlbom}
\thanks{These authors contributed equally to this work and are listed alphabetically.}
\affiliation{Neutron Scattering Division, Oak Ridge National Laboratory,
Oak Ridge, Tennessee 37831, USA}

\author{Daniel Kaplan}
\thanks{These authors contributed equally to this work and are listed alphabetically.}
\affiliation{Center for Materials Theory, Department of Physics and Astronomy,
Rutgers, The State University of New Jersey,
Piscataway, New Jersey 08854, USA}

\author{Premala Chandra}
\affiliation{Center for Materials Theory, Department of Physics and Astronomy,
Rutgers, The State University of New Jersey,
Piscataway, New Jersey 08854, USA}

\author{Cristian D. Batista}
\email{cbatist2@utk.edu}
\affiliation{Department of Physics and Astronomy,
The University of Tennessee, Knoxville, Tennessee 37996, USA}
\affiliation{Neutron Scattering Division and Shull-Wollan Center,
Oak Ridge National Laboratory, Oak Ridge, Tennessee 37831, USA}

\date{\today}
\begin{abstract}
In low density systems with strong interactions electrons are expected to crystallize into a Wigner solid. Recently, advances in two-dimensional systems where electrons carry Berry curvature have added a topological dimension to Wigner crystallization. Using a model for pseudospin interactions in a topological Wigner crystal, here we show that the competition between ferromagnetic Heisenberg exchange $J$ and the chiral interaction $\gamma$ on the triangular lattice stabilizes a zero-field skyrmion crystal whose density is continuously tunable through the ratio $\gamma/J$. The chiral interaction originates from finite Berry curvature in an underlying time-reversal broken Wigner crystal. In the continuum limit, the chiral interaction acts as a chemical potential for skyrmions, while higher-order gradient terms beyond the nonlinear sigma model select the skyrmion density. 
At large chiral coupling, we uncover a 24-site tetra-skyrmion crystal
carrying four units of topological charge per magnetic unit cell,
which becomes degenerate with four-sublattice tetrahedral order as
$J\to0$. We characterize the magnon structure for these phases and discuss the transition between the tunable skyrmion crystal to these states at large $\gamma/J$.  
Our results establish the phase diagram of ferromagnetic topological Wigner crystals.
\end{abstract}

\maketitle

\section{Introduction}
\label{sec:introduction}

A unifying framework for the interplay of strong interactions, electron
localization, and the spontaneous breaking of translation symmetry, all of
which are expected, for example, in the FQAH regime, is the Wigner crystal
(WC)~\cite{Wigner1934,TanatarCeperley1989}. This paradigm has recently been
extended to bands with nontrivial quantum geometry, giving rise to the
\emph{anomalous Hall crystal} (AHC): a translation-symmetry-breaking
electronic solid that spontaneously acquires a nonzero Chern
number~\cite{Dong_2024,Soejima2024,DongPatriStability2024,Zhou2024moireless,Tan2024,Sheng2024AHC,Zeng2024}.
This development sharpens a basic question: how are the properties of a
Wigner crystal altered once the underlying electronic bands carry Berry
curvature?

This is important in the context of a competing phase of matter now actively researched in low-dimensional, low density systems: the integer and fractional quantum anomalous Hall effects (IQAH and
FQAH). These states have now been
observed across several material platforms, from rhombohedral multilayer
graphene~\cite{Lu2024,Han2024,Xie2024} to twisted moir\'e transition-metal
dichalcogenides (TMDs)~\cite{Cai2023,Zeng2023,Park2023,Xu2023}, and have
been further resolved through local and magnetic
imaging~\cite{Redekop2024,Ji2024} as well as fractional quantum spin Hall
measurements~\cite{Kang2024}. 
What sets these states apart from the
conventional quantum Hall effect (QHE) is that the driver of the fractionalization is the Berry curvature of the underlying Bloch states rather
than an external magnetic field.

Equally important (and unlike the ordinary quantum Hall effect) these Bloch states are continuously
tunable, through the twist angle, the displacement field, or strain. The energetics in these materials suggest strong interactions dictate the ground
state~\cite{crepel2023anomalous,Reddy2023,ReddyFu2023,Wang2024FCI}, making them also hosts of Wigner crystallization \cite{Drummond2009}.
\begin{figure*}[t]
    \centering \includegraphics[width=0.95\textwidth]{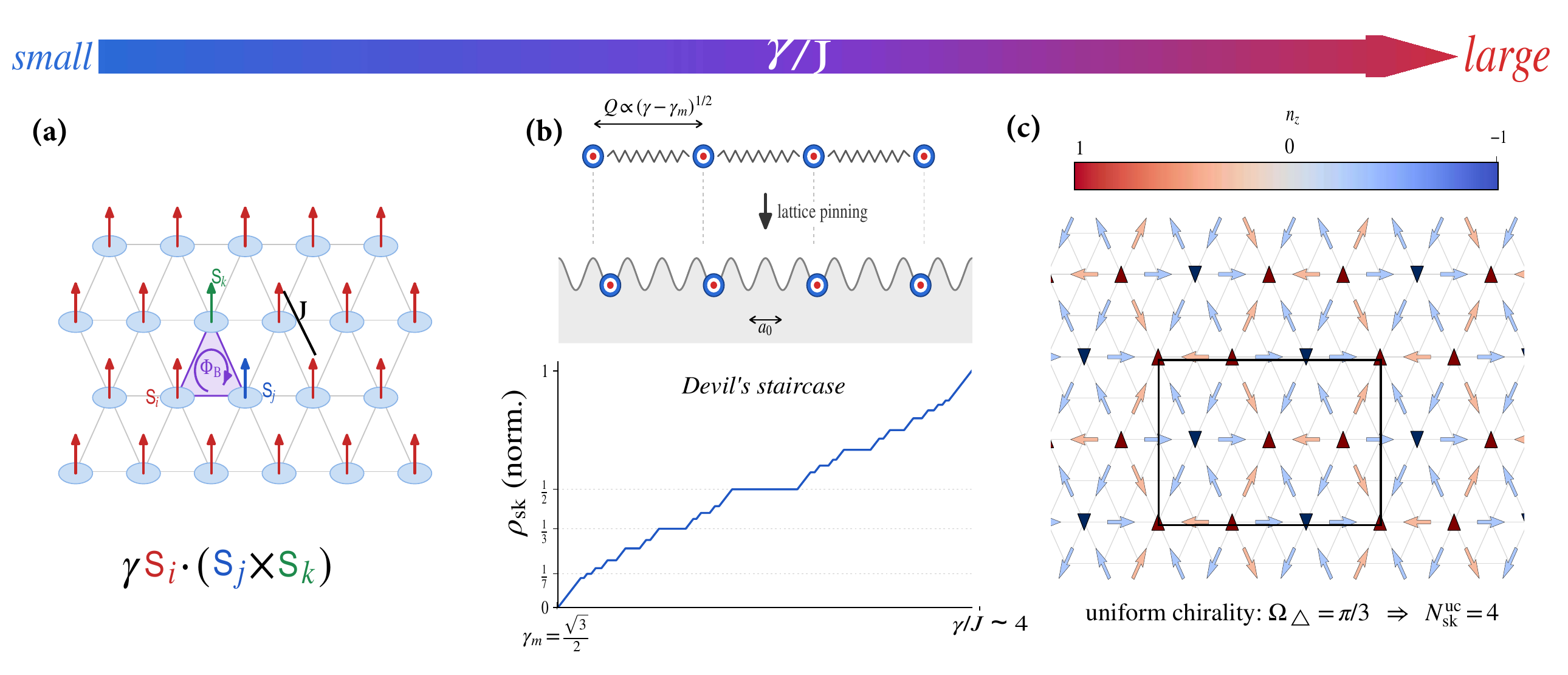}
\caption{Illustration of the topology-induced competition. (a) In the ferromagnetic Wigner crystal, the pseudospin degree of freedom is coupled ferromagnetically via a Heisenberg exchange $J$. Without any Berry flux, this state is fully polarized. In the presence of a finite Berry flux, three-particle exchange contributes an additional phase $\Phi_B$, which determines the chiral coupling $\gamma$. (b) As the chiral coupling is increased relative to $J$, the ferromagnet becomes unstable. At the critical value $\gamma/J = \sqrt{3}/2$, the system undergoes a set of commensurate-incommensurate transitions of different double-skyrmion $N=2$ crystals, giving a Devil's staircase pattern. (c) In the extreme limit when $J \to 0$, a new order emerges, and the ground state is a 24-site tetra-skyrmion crystal with $N=4$ skyrmion number. }
\label{fig:schematic}
\end{figure*}
The observation that the I- and F-QAH phases frequently appear in close proximity to
ordinary, lattice-commensurate flavor-polarized
states~\cite{Cai2023,Xu2023,ReddyFu2023}
reinforces the key role played by interactions; this proximity strongly suggests
that \emph{competition} between neighboring orders is responsible for the
rich phase diagrams observed in these materials. In the quantum Hall realm, strong interactions \cite{Sondhi93} lead to the emergence of skyrmions as the fundamental quasiparticles dictating the crossover from integer to fractional states. It is therefore worthwhile asking what role do topological textures play in Wigner crystals.

We answer this question by studying a minimal model for the stability of the Wigner
crystal in the presence of a competing, \emph{topological} interaction. We focus on ferromagnetic states, which are related to the flavor polarized IQAH seen in many material platforms.
We retain the two- and three-particle exchange processes that govern the most
relevant magnetic excitations of the
crystal~\cite{Thouless1965,Roger1983,Misguich1999,Bernu2001,Esterlis2025}.
In the presence of a finite Berry phase and broken time-reversal symmetry -- neither of which are found in the conventional Wigner crystal -- 
the three-particle ring-exchange process accomplishes two
things~\cite{SenChitra1995,KSKim2025}: it can invert the sign of the
two-spin Heisenberg interaction from antiferromagnetic to ferromagnetic, and,
crucially, it acquires an imaginary part that is precisely the scalar spin
chirality $\mathbf{S}_i\cdot(\mathbf{S}_j\times\mathbf{S}_k)$ of a
three-spin plaquette on the triangular lattice~\cite{Wen89}. 

We show, using complementary analytical and numerical methods, that the uniformly polarized ferromagnet is unstable toward a 
double skyrmion 
crystal~\cite{Muhlbauer2009,Yu10,nagaosa2013topological} at the critical ratio of the chiral coupling $\gamma$ to the Heisenberg exchange $J$,
$\gamma/J = \sqrt{3}/2$. A phase transition occurs in which the continuously
varying order parameter is the skyrmion density  $\rho_{\rm sk}$.
We assume neither a Dzyaloshinskii--Moriya interaction (DMI) nor any breaking
of the underlying lattice symmetry. Instead, the scalar chirality that
arises naturally in our model, and is governed by the Berry phase, acts as a ``chemical potential'' for chiral textures, driving a transition from ordinary ferromagnetism into a
skyrmion crystal similar to the one that has been conjectured for Chiral Stoner magnetism in Dirac bands~\cite{Dong2024}. 

Remarkably, $\rho_{\rm sk}$ grows continuously from zero and is
controlled by the ratio $\gamma/J$. It is only weakly sensitive to an
out-of-plane magnetic field, which induces a first-order transition back to
the uniform ferromagnet. As the chiral coupling increases further, the system passes through a series
of incommensurate states that fold the underlying triangular lattice.
We explain this through a
Frenkel--Kontorova mechanism, in which commensuration pinning to the lattice locks the skyrmion density onto a sequence of rational
plateaus, sometimes known as the ``Devil's staircase" ~\cite{Bak1982,Aubry1983,BraunKivshar1998}.

In the limit of very strong chiral coupling, we uncover a new spin phase
built from $24$ spins, a tetra-skymrion crystal (containing $N=4$ skyrmions). When the
Heisenberg exchange is tuned to $J=0$, this state becomes degenerate
with the four-sublattice tetrahedral order familiar from the $J>0$
antiferromagnetic model~\cite{Martin2008,Akagi2012}.

Finally, we provide directly accessible fingerprints of these phases through
the dynamical structure factor, and we discuss the transport signatures
that would reveal them~\cite{Taguchi01,Kurumaji2019,HayamiMotome2021}.
Our results offer a new understanding of the stability of Wigner crystals \cite{DongPatriStability2024,desrochers2026elastic,guo2025correlation,Soejima2025} and emergent chiral textures driven by the interplay of Berry curvature, lattice pinning and
electronic chirality. We thus unify frustrated magnetism and chiral magnetic textures with moir\'e matter and topological phases driven by Berry curvature.

The remainder of this paper is organized as follows. We begin in
Sec.~\ref{sec:summary} with a summary of our main results and the resulting
phase diagram. In Sec.~\ref{sec:origin_chiral_interaction} we show how the
effective chiral interaction emerges from a topological Wigner crystal. A
study of the effective spin Hamiltonian in the continuum limit is presented
in Sec.~\ref{sec:holomorphic_to_crystal}. The full phase diagram, bridging
the continuous topological phenomena and their discrete lattice realization,
is presented in Sec.~\ref{sec:Phase_Diagram}. In
Sec.~\ref{sec:experiment} we discuss experimental signatures, and we close
in Sec.~\ref{sec:discussion} with concluding remarks and an outlook for
future research directions.

\section{Summary of the Main Results}
\label{sec:summary}

\subsection{Minimal Model}

Motivated by these experimental developments and theoretical questions, we study the phase diagram of classical spins on the triangular lattice interacting through a ferromagnetic Heisenberg exchange and a scalar-chirality interaction generated by integrating out the electronic degrees of freedom. The minimal model is
\begin{equation}
H = -J\sum_{\langle ij\rangle}\mathbf{S}_i\cdot\mathbf{S}_j
-\gamma\sum_{\triangle_{ijk}}^{\circlearrowleft}
\mathbf{S}_i\cdot\left(\mathbf{S}_j\times\mathbf{S}_k\right)
-h\sum_i S_i^z ,
\label{eq:central}
\end{equation}
where the second sum runs over all elementary triangular plaquettes of the triangular lattice, including both upward- and downward-pointing triangles, with the sites $(i,j,k)$ ordered counterclockwise when viewed along the positive $z$ direction. The model describes the competition between the ferromagnetic two-spin exchange $J>0$ and the three-spin topological exchange $\gamma$, together with the effect of an external magnetic field $h$. In topological Wigner crystals, the scalar-chirality interaction originates from the Berry curvature of the electronic state. Consequently, $\gamma$ can be tuned through the carrier density, electrostatic gating, or layer engineering and may become comparable to, or even exceed, the bare exchange interaction.

\subsection{The Phase Diagram and Key Findings}

Our central result is that the competition between Heisenberg exchange ($J$), scalar chirality ($\gamma$) and lattice terms stabilizes zero-field tunable skyrmion states. 
We find that the chiral term, which acts as a topological chemical potential for skyrmion number, allows the skyrmion density to be continuously tuned through the ratio $\gamma/J$. 

To connect the universal predictions of continuum theory with the discrete microscopic reality of the triangular lattice, our combined analytic and computational approach bridges the full hierarchy of length scales:

\begin{itemize}

\item {\bf Long-Wavelength Regime:} The polarized state undergoes a continuous instability into a double-skyrmion crystal, where the ordering wave vector $Q$ emerges as the single scale governing both texture size and spacing.  
\item {\bf Field-Dependence:} The spin textured phase is only weakly distorted by an out-of-plane field until it undergoes a discontinuous transition to a fully polarized state.

\item{\bf Lattice-Pinning Domain:} As $Q$ increases, the skyrmion density locks into commensurate plateaus, a direct manifestation of Frenkel-Kontorova pinning.

\item{\bf Lattice-Scale Limit:} At extreme chiral coupling ($\gamma/J \ge 4$), the sequence terminates in a novel 24-sublattice tetra-skyrmion crystal. This phase realizes an unusual topological charge of $|N_{sk}| = 4$, demonstrating how discrete lattice geometry regularizes and enhances topological orders in the strong-coupling limit .
\item{\bf Dynamical Fingerprints:} We compute the dynamical spin structure factors to uniquely distinguish the competing ordered phases, revealing distinct  magnon spectra and momentum-resolved resonances that serve as clear roadmaps for neutron scattering experiments.

\end{itemize}

By bridging the continuum and lattice regimes, our results demonstrate that the underlying electronic topology acts as a primary driver of tunable magnetic order, offering a natural candidate for explaining the symmetry-broken cascades observed in fractional Chern insulators.

\section{Chiral Interaction in Topological Wigner Crystals}
\label{sec:origin_chiral_interaction}

Here we demonstrate how multi-particle exchange processes within a semiclassical Wigner crystal intrinsically generate both ferromagnetic coupling and scalar spin chirality.
The low-energy dynamics are generated by particle-exchange processes and have been derived for successive semiclassical approximations for moderate $r_s \sim 40$ \cite{Joy2025,kim2022interstitial,kim2024dynamical}. It is generally known that in the low density limit (large $r_s$), the resultant spin Hamiltonian is ferromagnetic \cite{Bernu2001}. The effect of defects and interstitials on the spin configuration has also been discussed lately \cite{kim2022interstitial}, showing the emergence of ferromagnetism at lower $r_s$ and at higher temperatures. 

Thus, we are motivated to consider the following scenario: we take spin-$1/2$ electrons on a triangular Wigner lattice, for which the leading two- and three-particle exchanges may be written as
\begin{equation}
    H_{\rm ex}
    =
    J_2
    \sum_{\langle ij\rangle}
    \left(
        P_{ij}+P_{ji}
    \right)
    -
    J_3
    \sum_{\triangle_{ijk}}^{\circlearrowleft}
    \left(
        e^{i\varphi_{ijk}}P_{ijk}
        +
        e^{-i\varphi_{ijk}}P_{ikj}
    \right).
    \label{eq:wigner_exchange_hamiltonian}
\end{equation}
Here, the second sum runs over all elementary triangular plaquettes, including both upward- and downward-pointing triangles, with the sites $(i,j,k)$  ordered counterclockwise when viewed along the positive $z$ direction. The operator $P_{ij}$ exchanges the spins on sites $i$ and $j$,
whereas
\(
P_{ijk}\equiv P_{ij}P_{jk}
\)
denotes the counterclockwise cyclic permutation
$i\to j\to k\to i$. Its inverse,
$P_{ikj}=P_{ijk}^{-1}$, denotes the corresponding clockwise
permutation.
The crucial ingredient is phase $\varphi_{ijk}$, 
which is the Berry phase associated with the corresponding exchange path.

For spin-$1/2$, the exchange operators obey
\begin{equation}
    P_{ij}
    =
    2\mathbf S_i\cdot\mathbf S_j
    +
    \frac{1}{2},
    \label{eq:two_spin_permutation}
\end{equation}
and
\begin{align}
    P_{ijk}+P_{ikj}
    &=
    P_{ij}+P_{jk}+P_{ki}-1,
    \label{eq:three_cycle_symmetric}
    \\
    P_{ijk}-P_{ikj}
    &=
    -4i\,
    \mathbf S_i\cdot
    \left(
        \mathbf S_j\times \mathbf S_k
    \right).
    \label{eq:three_cycle_antisymmetric}
\end{align}
Substituting Eqs.~\eqref{eq:three_cycle_symmetric} and
\eqref{eq:three_cycle_antisymmetric} into
Eq.~\eqref{eq:wigner_exchange_hamiltonian} gives
\begin{align}
    e^{i\varphi_{ijk}}P_{ijk}
    +
    e^{-i\varphi_{ijk}}P_{ikj}
    &=
    \cos\varphi_{ijk}
    \left(
        P_{ij}+P_{jk}+P_{ki}-1
    \right)
    \nonumber\\
    &\quad
    +
    4\sin\varphi_{ijk}\,
    \mathbf S_i\cdot
    \left(
        \mathbf S_j\times \mathbf S_k
    \right).
    \label{eq:wigner_three_exchange_decomposition}
\end{align}
Thus three-particle exchange produces both a symmetric exchange
renormalization and a scalar-chirality interaction:
\begin{equation}
    H_{\rm ex}
    =
    -
    \sum_{\langle ij\rangle}
    J^{(W)}_{ij}\,
    \mathbf S_i\cdot\mathbf S_j
    -
    \sum_{\triangle_{ijk}}^{\circlearrowleft}
    \gamma^{(W)}_{ijk}\,
    \mathbf S_i\cdot
    \left(
        \mathbf S_j\times \mathbf S_k
    \right)
    \label{eq:wigner_effective_spin_model}
\end{equation}
with
\begin{equation}
    \gamma^{(W)}_{ijk}
    =
    4J_3\sin\varphi_{ijk}
    \label{eq:wigner_gamma}
\end{equation}
where a constant term has been absorbed.
The effective pair exchange $J^{(W)}_{ij}$ receives contributions from
both $J_2$ and the $\cos\varphi_{ijk}$ part of the three-particle
exchange.  Its precise value depends on the counting of exchange loops
sharing the bond $ij$, but schematically
\begin{equation}
    J^{(W)}_{ij}
    =
    -4J_2
    +
    2J_3
    \sum_{\triangle \in ij}
    \cos\varphi_{\triangle}.
    \label{eq:wigner_pair_exchange_schematic}
\end{equation}
Hence the net nearest-neighbor exchange can become ferromagnetic even
when the bare two-particle exchange is antiferromagnetic, for example
in the presence of defects or competing exchange paths.

Eq.~\eqref{eq:wigner_gamma} shows that the chiral interaction is
present only when the exchange path carries a time-reversal-odd phase:
$\sin\varphi_{ijk}\neq0$.  For $\varphi_{ijk}=0$ or $\pi$, the
three-particle exchange still renormalizes the symmetric spin exchange,
but it does not generate scalar chirality.  Therefore a Wigner crystal
with ferromagnetic effective exchange and nontrivial Berry phase maps
onto the same spin model as Eq.~\eqref{eq:central}. This is the setting relevant to
chiral Wigner and anomalous Hall crystals formed from Chern
bands, where the orbital Berry curvature supplies the phase
$\varphi_{ijk}$.  

While these recent experiments motivate our specific focus on the strongly localized Wigner regime, the emergence of this chiral spin coupling is a more universal phenomenon. Identical effective spin dynamics, governed by the competition between $J$ and $\gamma$, emerge from both Hubbard and Kondo lattice models at partial filling when the interacting electronic bands possess intrinsic momentum-space Berry curvature; these derivations can be found in Appendix~\ref{app:kondohubbard}.
In these metallic or semi-metallic models, integrating out the itinerant electrons in the presence of broken time-reversal symmetry yields a multi-spin exchange where the scalar chirality coupling is directly proportional to the gauge-invariant  Berry phase piercing the elementary triangular plaquettes. Consequently, the fundamental mechanism described by Eq.~\eqref{eq:central} whereby intrinsic electronic topology acts as the active driver of tunable chiral magnetic textures should be applicable across a broad spectrum of strongly correlated topological materials. 

In the following sections, we analyze the phase diagram of this effective model by tracking its evolution from the macroscopic continuum down to the discrete lattice scale. We start with the long-wavelength theory in Sec.~\ref{sec:holomorphic_to_crystal}, demonstrating how the competition between the topological chiral interaction and microscopic lattice gradients uniquely determines the equilibrium length scale of the skyrmion crystal.


\section{Long Wavelength Theory}
\label{sec:holomorphic_to_crystal}

In the long-wavelength limit, the semiclassical description becomes asymptotically controlled even at fixed $S$. The uniform ferromagnetic state at $Q=0$ is exact, while the interaction vertices generated by a slowly varying noncollinear texture are derivative-coupled and therefore vanish as $Q\to0$. Quantum corrections are consequently suppressed by powers of the characteristic wave vector $Q$, measured in inverse lattice units, allowing the conventional $1/S$ expansion to be reorganized as a controlled small-$Q$ expansion. We therefore begin with the classical limit, which provides the
leading term in this controlled small-$Q$ expansion, and later
characterize fluctuations about the resulting ordered phases using
low-temperature Landau--Lifshitz dynamics. In the harmonic regime,
the linearized Landau--Lifshitz equations yield the same excitation
frequencies as linear spin-wave theory, while the classical dynamical
structure factor is converted into its quantum counterpart using the
standard quantum--classical correspondence factor \cite{schofield1960space, zhang2019dynamical, dahlbom2024quantum}.

To connect this classical description with the microscopic quantum
Hamiltonian derived in Sec.~III, we write the quantum spin operator as
$\hat{\mathbf S}_i=S\mathbf n_i$, where $\mathbf n_i$ is a classical
unit vector. The two-spin and three-spin interactions then acquire
factors of $S^2$ and $S^3$, respectively. We absorb these factors into
the definitions of the classical couplings,
\begin{equation}
J\equiv J_{\rm q}S^2,
\qquad
\gamma\equiv\gamma_{\rm q}S^3,
\qquad
h\equiv h_{\rm q}S.
\end{equation}
Thus, throughout the following classical analysis, $J$, $\gamma$, and
$h$ denote the couplings of the unit-vector model in
Eq.~\eqref{eq:central}. Hereafter, all energies are measured in units of $J$, so that the
numerical values of $\gamma$ and $h$ denote the dimensionless ratios
$\gamma/J$ and $h/J$, respectively.

Our starting point is the two-dimensional \(O(3)\) nonlinear sigma model for a unit vector field
\(\mathbf n(\mathbf r)=(n_x,n_y,n_z)\), \(\mathbf n^2=1\):
\begin{equation}
\mathcal H_{\sigma}
=
\frac{\rho_s}{2}\int d^2r\, (\partial_\mu \mathbf n)\cdot(\partial_\mu \mathbf n),
\label{eq:Hsigma}
\end{equation}
where repeated spatial indices \(\mu=x,y\) are summed over.  This continuum theory is scale invariant in two dimensions and supports topologically nontrivial textures classified by the skyrmion number
\begin{equation}
N_{\rm sk}
=
\frac{1}{4\pi}\int d^2r\,
\mathbf n\cdot
\left(
\partial_x \mathbf n \times \partial_y \mathbf n
\right).
\label{eq:Qdef}
\end{equation}

A central result of the model is the Belavin-Polyakov bound ~\cite{Belavin1975},
\begin{equation}
\frac{1}{2}\,(\partial_\mu \mathbf n)^2
\ge
\left|
\mathbf n\cdot(\partial_x \mathbf n\times \partial_y \mathbf n)
\right|,
\label{eq:BPbound}
\end{equation}
which implies
\begin{equation}
E_\sigma \ge 4\pi \rho_s |N_{\rm sk}|.
\label{eq:BPenergy}
\end{equation}

Configurations that saturate this bound form a continuous,
scale-degenerate family of skyrmion solutions. The bound is saturated
by self-dual configurations obeying
\begin{equation}
\partial_\mu \mathbf n
=
\pm \epsilon_{\mu\nu}\,
\mathbf n\times \partial_\nu \mathbf n,
\label{eq:selfdual}
\end{equation}
and that a convenient way of exhibiting these solutions is to use the stereographic coordinate
\begin{equation}
w(\mathbf r)
=
\frac{2(n_x+i n_y)}{1-n_z},
\label{eq:stereographic}
\end{equation}
which maps the unit sphere onto the complex plane.  In terms of \(w\), the sigma-model energy becomes
\begin{equation}
\mathcal H_\sigma
=
\rho_s \int d^2r\,
\frac{
\partial_z w\,\partial_{\bar z}\bar w
+
\partial_{\bar z}w\,\partial_z\bar w
}{
\left(1+|w|^2/4\right)^2
}.
\label{eq:Hsigma_w}
\end{equation}
with \(z=x+iy\), \(\partial_z=\frac12(\partial_x-i\partial_y)\), and
\(\partial_{\bar z}=\frac12(\partial_x+i\partial_y)\).  Equation~\eqref{eq:selfdual} then reduces to the Cauchy-Riemann condition, so any holomorphic or antiholomorphic function generates an exact skyrmion solution. 
In particular, this means a single skyrmion centered at \(z_0\) is represented by
\begin{equation}
w(z)=\frac{\lambda}{z-z_0},
\label{eq:single_sk}
\end{equation}
where $\lambda = 1/Q$ or, more generally, by rational maps for higher charge. 
Thus, any holomorphic function in principle represents a valid solution. 

In order to favor configurations with nonzero skyrmion density, one may add to the continuum functional a term proportional to the topological density,
\begin{equation}
\mathcal H
=
\mathcal H_\sigma
-
\mu N_{\rm sk} 
\label{eq:Hcont_mu}
\end{equation}
which has the same form as the continuum limit 
of Eq.~\eqref{eq:central} with $\mu = 8 \pi \gamma$.

Equation~\eqref{eq:Hcont_mu} makes the instability transparent. For the skyrmion chirality favored by $\mu$, configurations that saturate the Belavin--Polyakov bound have the energy,
\begin{align}
E= \left(4\pi\rho_s-\mu\right)N_{\rm sk}.
\end{align}
For $\mu<\mu_c=4\pi\rho_s$, the coefficient is positive and the ground state has zero skyrmion density. At $\mu=\mu_c$, configurations with arbitrary skyrmion number and size become degenerate. For $\mu>\mu_c$, however, the energy is lowered by increasing $N_{\rm sk}$ and hence the skyrmion density, $\rho_{\rm sk}\propto Q^2$. Because the nonlinear sigma model contains no scale-setting term, minimization drives $Q\to\infty$, corresponding to a direct instability from zero to formally infinite skyrmion density. Thus, while the tendency toward nonzero topological density is already present in the nonlinear sigma model, a stable finite skyrmion size can arise only from terms that break its scale invariance. On the lattice, these are the higher-gradient corrections neglected in Eq.~\eqref{eq:Hsigma}. Phenomenologically, they extend the energy density from $e(Q)=aQ^2$ to
\begin{equation}
e(Q)
=
a Q^2
+
b Q^4
+
c Q^6
+ \mathcal{O}(Q^8).
\label{eq:e_lambda}
\end{equation}
where the coefficient $b$ provides the leading scale-setting contribution.

Since the starting point of Eq.~\eqref{eq:central} is the Wigner crystal, we will work specifically on a regular triangular lattice where this logic follows naturally. The long-wavelength expansion of Eq.~\eqref{eq:central} contains the sigma-model functional as its leading term, but it also generates contributions with four and six spatial derivatives.  Those  are the first terms that lift the scale degeneracy of the holomorphic skyrmions and therefore control the density of the skyrmion crystal. 

\subsection{Gradient Expansion on the Triangular Lattice}

The microscopic triangular lattice of the Wigner crystal is generated by the primitive vectors
\begin{equation}
\mathbf a_1=a_0(1,0),
\qquad
\mathbf a_2=a_0\left(-\frac12,\frac{\sqrt3}{2}\right).
\label{eq:triangular_vectors}
\end{equation}
Here we will assume that the lattice constant $a_0=1$, leading to a unit-cell area of
\begin{equation}
A_{\rm uc}=\frac{\sqrt3}{2}.
\label{eq:Auc}
\end{equation}
To derive the continuum theory, we expand the spin field about each lattice site,
\begin{equation}
\mathbf n(\mathbf r+\boldsymbol\delta)
=
\sum_{m=0}^{\infty}
\frac{1}{m!}
(\boldsymbol\delta\!\cdot\!\nabla)^m \mathbf n.
\label{eq:Taylor_n}
\end{equation}
The Heisenberg term then yields, after summing over the six nearest-neighbor bonds and converting the lattice sum into an integral,
\begin{equation}
\mathcal H_{\rm Heis}
=
\int d^2r\,
\left[
\frac{\kappa_2}{2}\,(\partial_\mu \mathbf n)^2
+
\kappa_4\,(\nabla^2\mathbf n)^2
+
\kappa_6\,\mathcal O_6^{(h)}
+\cdots
\right],
\label{eq:Heis_grad_main}
\end{equation}
where \(\mathcal O_6^{(h)}\) denotes the symmetry-allowed sixth-order gradient invariants.  For the triangular lattice, the explicit derivation given in Appendix~\ref{app:heisenberg} yields
\begin{equation}
\mathcal H_{\rm Heis}
\simeq
\frac{3J}{4A_{\rm uc}}\int d^2r\, (\partial_\mu \mathbf n)^2
-\frac{3J}{64A_{\rm uc}}\int d^2r\, (\nabla^2 \mathbf n)^2.
\label{eq:Heis_grad_explicit}
\end{equation}
The first term is the continuum spin stiffness. The second is the leading four-derivative correction generated by the lattice. Because its contribution to the energy density scales as $Q^4$ for a skyrmion texture with characteristic wavevector $Q$, it provides the dominant mechanism that selects the equilibrium skyrmion density once the quadratic coefficient changes sign.

The chiral term admits an analogous expansion. At leading order it reduces to the topological density (see Appendix~\ref{app:chiral}),
\begin{equation}
\begin{aligned}
\mathcal H_{\chi}
\simeq{}&
-2\gamma \int d^2r\,
\mathbf n\!\cdot\!
\bigl(\partial_x \mathbf n\times \partial_y \mathbf n\bigr)
\\[2pt]
&-\frac{\gamma\sqrt{3}}{16A_{\rm uc}}
\int d^2r\, \mathbf n\!\cdot\!\Big[
\partial_x \mathbf n\times \partial_y^3 \mathbf n
+\partial_x \mathbf n\times \partial_x^2\partial_y \mathbf n
\\
&\hspace{2.8cm}
+\partial_x\partial_y \mathbf n\times \partial_y^2 \mathbf n
+\partial_x\partial_y^2 \mathbf n\times \partial_y \mathbf n
\\
&\hspace{2.8cm}
+\partial_x^2 \mathbf n\times \partial_x\partial_y \mathbf n
+\partial_x^3 \mathbf n\times \partial_y \mathbf n
\Big].
\end{aligned}
\label{eq:Chiral_grad_main}
\end{equation}
This contribution is equally important because it also enters at order $Q^4$, together with the four-derivative Heisenberg correction. Consequently, both microscopic interactions cooperate in determining the optimal skyrmion density.

\subsection{Energy Expansion in Powers of $Q$}

The long-wavelength theory becomes especially transparent when evaluated on a skyrmion crystal characterized by a single emergent wavevector $Q$; it simultaneously controls the skyrmion size and the inter-skyrmion spacing. 
Equivalently, $Q$ may be viewed as the inverse lattice constant of an ideal skyrmion crystal or, more generally, as the  magnitude of ordering wave vector of a triple-\(Q\) texture. From the following relations,
\begin{equation}
\partial_\mu \mathbf n \sim Q,
\qquad
\partial_\mu\partial_\nu \mathbf n \sim Q^2,
\qquad
\mathbf n\cdot(\partial_x \mathbf n\times \partial_y \mathbf n)\sim Q^2,
\label{eq:scaling_derivatives}
\end{equation}
it follows that the gradient expansion generates an energy density of the form \eqref{eq:e_lambda}.
The physical meaning is immediate:

(i) the coefficient \(a\) controls the instability of the fully polarized state toward a state with nonzero skyrmion density;

(ii) the coefficient \(b\) is the first one that breaks the continuum scale invariance and therefore controls the optimal density once \(a\) changes sign;

(iii) the coefficient \(c\) stabilizes the short-distance behavior and becomes important when \(b<0\) or close to a first-order transition. 

More explicitly, the coefficient \(a\) receives contributions only from the leading two-derivative Heisenberg and the leading topological chiral 
terms :
\begin{equation}
a(\gamma)=a_h+\gamma a_\chi
= a_\chi(\gamma-\gamma_m)
\label{eq:agamma}
\end{equation}
where here $\gamma$ and $\gamma_m$ are in units of $J$.
For a holomorphic skyrmion crystal ansatz, both pieces can be obtained analytically because the Heisenberg energy saturates the Belavin-Polyakov bound while the chiral term is proportional to the skyrmion density.  This gives the critical value at which $a(\gamma)$, the $Q^2$ coefficient, changes sign to be
\begin{equation}
\gamma_m 
=-\frac{a_h}{a_\chi}=\frac{\sqrt3}{2}. 
\label{eq:gamma_m}
\end{equation}

However, this condition alone does not fix the equilibrium $Q$: once \(a\) becomes negative, the minimizing value of $Q$ is determined by the balance between the destabilizing $aQ^2$ and the stabilizing $bQ^4$ and $cQ^6$ terms.  In particular, when \(b>0\) the transition is continuous and the skyrmion wavevector grows smoothly from zero; when \(b<0\), the sixth-order term is needed to stabilize the energy and a first-order transition becomes possible.
The coefficients $b$ and $c$ are obtained from a numerical evaluation of the lattice energy functional for the skyrmion crystal configurations described in the following section. (See Appendix~\ref{sec:app_coeff_fitting} for details). The results show that $b$ is positive in the vicinity of the transition, implying that the instability of the fully polarized state is continuous and the skyrmion density grows smoothly from zero.

\subsection{Structure of the Ordered State \label{sec:dlb_skx_ansatz}}

\begin{figure}[htbp]
    \centering
    \includegraphics[width=\textwidth,height=0.40\textheight,keepaspectratio]{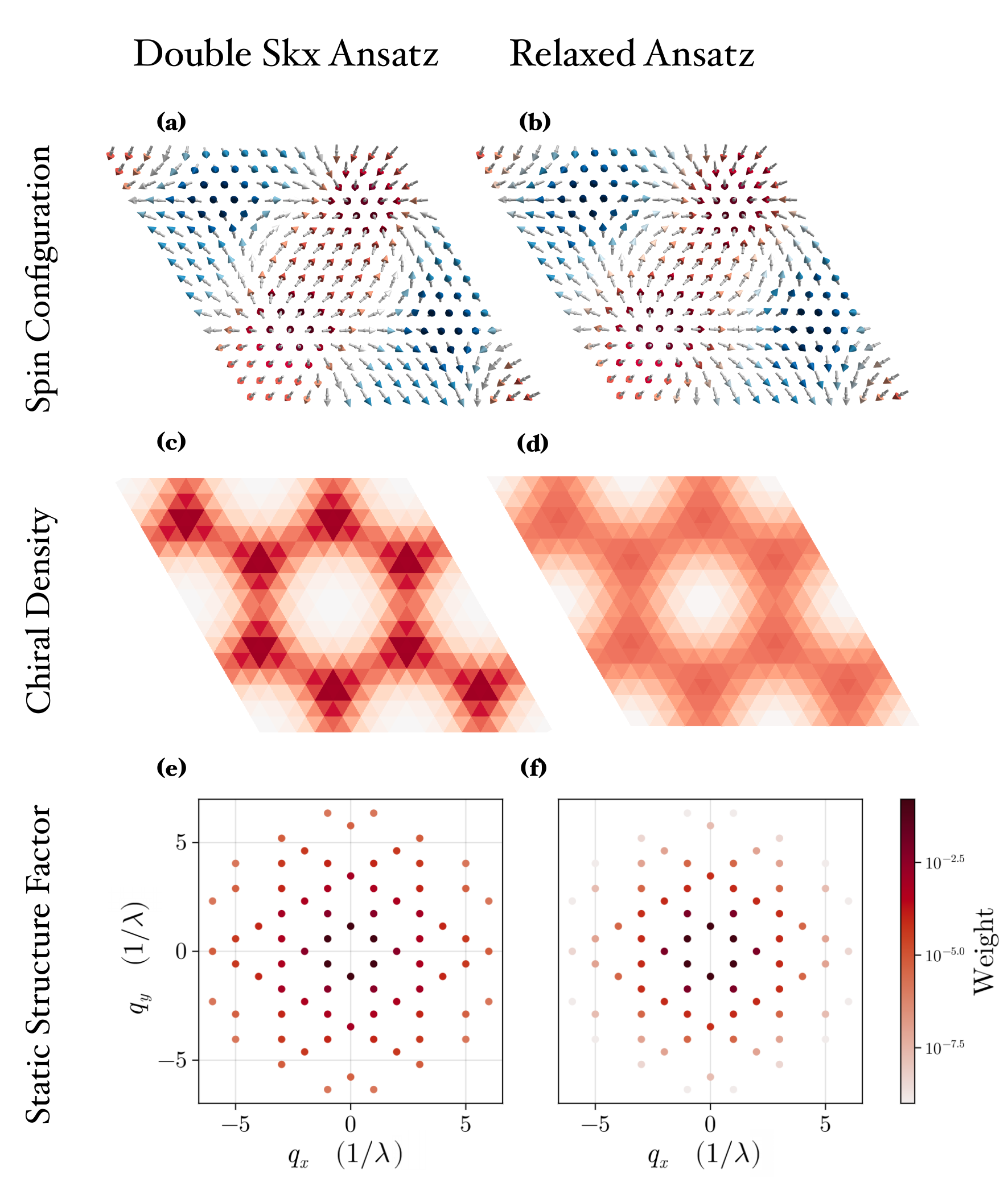}
    \caption{%
    Comparison between the triple-$\mathbf{Q}$ ansatz of Eq.~\eqref{eq:tripleQansatz} and the numerically relaxed configuration ((a) and (b)) with 
    $Q = 0.0625$ and $\gamma=0.93$. The spatial distribution of the skyrmion density is largely preserved under relaxation, with only
    quantitative distortions of the profile ((c) and (d)), corresponding to a transfer of weight to lower-order Fourier components ((e) and (f), colors in log scale). The Brillouin zone boundary lies outside of the panel boundaries of (e) and (f); without log color scaling, weight would only be visible on the six innermost components in both cases.
    }
    \label{fig:relaxation}
\end{figure}

The discussion above does not require the explicit form of the crystal solution, only that it is characterized by a single wavevector $Q$. Nevertheless, it is useful to introduce a minimal variational ansatz that captures the essential structure of the ordered phase. A useful guiding principle is that a skyrmion crystal with zero net magnetization must contain equal populations of textures whose cores have opposite magnetization, while preserving a uniform sign of the scalar chirality. This immediately suggests a structure composed of two types of skyrmions with identical topological charge but opposite core polarization, i.e., a double-skyrmion crystal.

A natural realization of such a state on the triangular lattice is provided by a triple-$\mathbf Q$ superposition~\cite{Martin2008,Ozawa2017},
\begin{equation}
\begin{split}
\mathbf n(\mathbf r)
&=
\frac{
\left(
\cos(2 \pi\mathbf Q_1\cdot \mathbf r),
\cos(2 \pi \mathbf  Q_2\cdot \mathbf r),
\cos(2 \pi \mathbf Q_3\cdot \mathbf r)
\right)
}{
\sqrt{
\cos^2(2 \pi \mathbf Q_1 \cdot \mathbf r)
+
\cos^2(2 \pi \mathbf Q_2\cdot \mathbf r)
+
\cos^2(2 \pi \mathbf Q_3\cdot \mathbf r)
}
},
\end{split}
\label{eq:tripleQansatz}
\end{equation}
with 
$\mathbf Q_{\nu} = 2\mathbf M_{\nu}/\lambda$ ($\nu=1,2,3$) and
\begin{equation}
\mathbf M_1=\frac{\mathbf b_1}{2},\qquad
\mathbf M_2=\frac{\mathbf b_2}{2},\qquad
\mathbf M_3=- \frac{(\mathbf b_1 + \mathbf b_2)}{2}.
\label{eq:Mpoints}
\end{equation}
in terms of the reciprocal lattice vectors $\mathbf b_1$ and $\mathbf b_2$ ($
\mathbf a_\mu \cdot \mathbf b_\nu = \delta_{\mu\nu}$). The wave vectors 
$M_{\nu}$ are half of reciprocal lattice vectors located at the midpoints of three edges of the hexagonal Brillouin zone that differ from each other by $\pm 120^{\circ}$ rotations. This construction represents the simplest nontrivial superposition of symmetry-related wave vectors compatible with the triangular lattice and naturally generates a periodic array of skyrmions with opposite core magnetization.

While Eq.~(\ref{eq:tripleQansatz}) is introduced here as a minimal variational construction motivated by symmetry and the constraint of vanishing net magnetization, similar double-skyrmion crystal structures have appeared in different physical contexts. In particular, a skyrmion crystal with topological charge $|N_{\rm sk}|=2$ was reported in the Kondo lattice model, where it emerges from the interplay between itinerant electrons and localized moments mediated by RKKY-type interactions and Fermi-surface effects \cite{Ozawa2017}. Likewise, the four-sublattice noncoplanar (tetrahedral) spin ordering reported in Ref.~\cite{Martin2008} can be interpreted as a dense limit of a double-skyrmion crystal.

It is important to emphasize that the mechanism discussed here is fundamentally different. In the present case, the double-skyrmion structure arises from a long-wavelength instability of the fully polarized state, driven by the competition between a topological term and lattice gradient corrections. This leads to a controlled expansion in powers of the characteristic wave vector $Q$, which continuously tunes the skyrmion density near the transition. By contrast, the skyrmion crystals in Refs.~\cite{Ozawa2017,Martin2008} are stabilized by microscopic electronic mechanisms that select multiple-$\mathbf Q$ states at a \textit{fixed} and finite wave vector magnitude $Q$. As a consequence, the corresponding skyrmion density is not a tunable quantity, but is instead set by microscopic electronic scales.

Although Eq.~\eqref{eq:tripleQansatz} is not an exact solution of the microscopic Hamiltonian, it provides a minimal ansatz that captures the essential qualitative features of the skyrmion crystal phase. In particular, it generates a periodic array of double skyrmions, with a skyrmion density, $\rho \sim |\mathbf Q_\nu|^2$, that is concentrated near the centers of the magnetic unit cell and suppressed in the interstitial regions. 

 Importantly, this ansatz is characterized by a single length scale, $Q^{-1}$, which simultaneously controls the skyrmion size and the spacing between skyrmions. The wave vector $Q$ is treated as a variational parameter and is determined by minimizing the energy.
To assess the accuracy of the ansatz, one can numerically relax the spin configuration starting from Eq.~\eqref{eq:tripleQansatz}. The resulting optimized configuration exhibits a very similar spatial distribution to the skyrmion density of the original ansatz, with only quantitative distortions of the profile. In particular, the location and topology of the double-skyrmion cores are preserved, while a reduction in higher harmonics primarily acts to sharpen the texture without altering its qualitative structure. 
This comparison, illustrated in Fig.~\ref{fig:relaxation}, demonstrates that the triple-$\mathbf Q$ ansatz provides an accurate and physically transparent representation of the skyrmion crystal phase.

The coefficients in the expansion Eq.~\eqref{eq:e_lambda} can be obtained numerically by computing the energy of the relaxed double-skyrmion crystal as a function of its characteristic wavevector $Q$, and fitting the resulting dependence to Eq.~\eqref{eq:e_lambda}.
As discussed in Appendix~\ref{sec:app_coeff_fitting}, the fitted values give $b>0$ in
the vicinity of $\gamma_m$, confirming that the zero-field instability of the fully polarized state is continuous. Furthermore, the coefficients found for $a_{\rm h}$ and $a_{\chi}$ saturated the theoretical bounds given by the nonlinear sigma model and predicting a transition at $\gamma_m=0.866$, very near to the analytical value of $\sqrt{3}/2$ given in Eq.~\eqref{eq:gamma_m}.

Near \(\gamma_m\), the transition into the fully polarized ferromagnetic state is continuous, and the energy-density expansion in Eq.~\eqref{eq:e_lambda} can be expressed as an analytic expansion in the skyrmion density \(\rho_{\rm sk} \sim Q^2\). Since the skyrmions themselves form a triangular lattice on top of the microscopic lattice, the area of the unit cell associated with the wavelength $\lambda = 1/Q$ is $A=\sqrt{3}\lambda^2/2$. For a double-skyrmion crystal, the corresponding skyrmion density is
$\rho_{\rm sk}= \tfrac{2}{A}=\frac{4 Q^2}{\sqrt{3}}$. We therefore recast Eq.~\eqref{eq:e_lambda} in terms of the density $\rho_{sk}$ as
\begin{equation}
e(\rho_{\rm sk})=\frac{\sqrt{3}}{4} a(\gamma)\rho_{\rm sk}+\frac{3}{16}b\rho_{\rm sk}^2+\frac{3\sqrt{3}}{64}c\rho_{\rm sk}^3+\mathcal{O}(\rho_{sk}^4) ,
\label{eq:exp}
\end{equation}
where $b>0$ reflects the effective repulsion between skyrmions.
We recall from Eq.~\eqref{eq:agamma} that the
coefficient $a(\gamma)$ changes sign at $\gamma_m$.
Minimizing the energy at zero field gives
\begin{equation}
\rho_{\rm sk}^*= - \frac{2}{\sqrt{3}}\frac{a}{b}+O(a^2),
\qquad
e^*=-\frac{a^2}{4b}+O(a^3).
\label{eq:est}
\end{equation}
 Equivalently, using $\rho_{\rm sk}=4Q^2/\sqrt{3}$ in Eq.~(32) gives
\begin{equation}
  Q^*(\gamma)
  =
  \sqrt{\frac{|a_\chi|}{2b_m}
  } \left(\gamma-\gamma_m\right)^{1/2}
  + {\cal O} \!\left[(\gamma-\gamma_m)^{3/2}\right],
  \label{eq:Qgamma}
\end{equation}
where $\gamma_m=\frac{\sqrt{3}}{2}$ and  $b_m\equiv b(\gamma_m)$. Corrections arising from the
$\gamma$ dependence of $b$ enter only at higher order.

The ordering wave vector therefore appears initially as a square root, with amplitude fixed by the single non-universal
coefficient $b_m$. In contrast, the quadratic coefficients $a_h$ and
$a_\chi$ are pinned by the Belavin--Polyakov bounds [Eq.~(28)] and are
independent of the texture; we can therefore conclude without reference to any fitted parameters that, at onset, $Q_*^2$ is linear in $\gamma$ with a crossing at $\gamma_m$. The precise form of $Q^*(\gamma)$, using a value of $b_m$ fit to the double skyrmion ansatz, is shown in Fig.~\ref{fig:shortwavelength_pd}~(b), where it is compared against unbiased numerical calculations.

To determine the phase boundary in a magnetic field, we now include the
Zeeman coupling to the spin texture. This term is given by
\begin{equation}
H_Z = - h m \sum_j n_j^z ,
\label{eq:Zee}
\end{equation}
which, in the continuum limit, becomes
\begin{equation}
H_Z = - \frac{h m}{A_{\rm uc}} \int d^2 r\, n^z(\mathbf r).
\end{equation}

To determine the phase boundary in a magnetic field, one must account for the fact that the Zeeman term does not primarily act by changing the skyrmion density. Numerically, the dominant effect of a weak out-of-plane field is instead to deform the internal shape of the skyrmion texture: the cores whose spins are antiparallel to the field contract in order to gain Zeeman energy, while the characteristic ordering wave vector $Q$ remains nearly unchanged over a broad field window. In other words, the field mainly redistributes the magnetization within a magnetic unit cell rather than changing the density of topological charge.

A simple estimate of the saturation field can be obtained within a rigid-crystal approximation, in which the skyrmion crystal retains its zero-field characteristic wave vector and hence its skyrmion density. The field then primarily lowers the energy of the fully polarized state through the Zeeman coupling. Using $M_{\rm SkX}^{(0)}=0$, the saturation field follows from the level-crossing condition between the skyrmion crystal and the fully polarized state,
\begin{equation}
\Delta e_0+\frac{m h_{\rm sat}}{A_{\rm uc}} \approx 0,
\end{equation}
where $\Delta e_0=e_{\rm SkX}(0)-e_{\rm FP}(0)<0$ is the zero-field condensation energy density. This immediately gives
\begin{equation}
h_{\rm sat} \approx \frac{A_{\rm uc}}{m}\,|\Delta e_0|.
\label{eq:hsat}
\end{equation}

By combining Eqs.~\eqref{eq:agamma}, ~\eqref{eq:est} and ~\eqref{eq:hsat}, we obtain
\begin{equation}
h_{\rm sat} \approx
 \frac{A_{\rm uc}}{4mb}\, a_\chi^2(\gamma-\gamma_m)^2
+O\!\left[(\gamma-\gamma_m)^3\right]
\label{eq:sat2}
\end{equation}
which indicates that  the field required to suppress the skyrmion crystal scales quadratically with the system's distance from the critical point  $\gamma=\gamma_m$.
Because the field-induced saturation transition point is defined by the crossing of the
total energies of two competing states, rather than the order parameter ($\rho_{sk}$) evolving smoothly to zero as a function of $h$, the mathematical framework inherently describes a discontinuous, first-order transition.

\begin{figure}[htbp]
    \centering
    \includegraphics[width=0.4\textwidth,keepaspectratio]{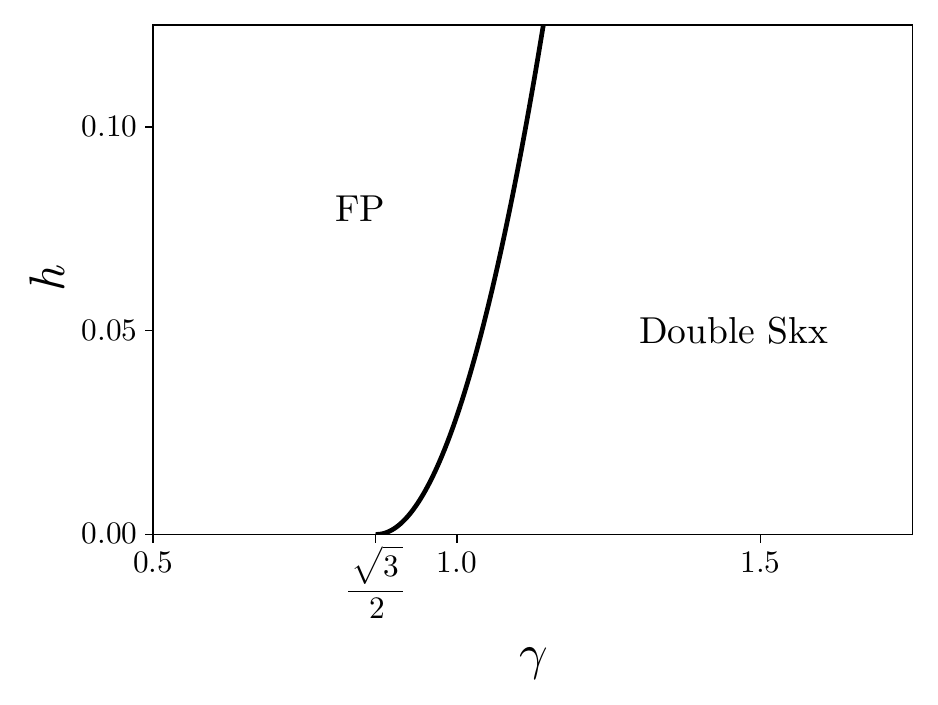}
    \caption{
    Long-wavelength phase diagram for the Hamiltonian of Eq.~\eqref{eq:central} including lattice corrections up to quartic order in inverse skyrmion lengthscale $1/\lambda$. ``FP'' is the fully polarized phase, and ``Double Skx'' is a crystal of double skyrmions. The spin texture of the Double Skx phase is illustrated in Fig.~\ref{fig:relaxation}.
    }
\label{fig:longwavelength_pd}
\end{figure}

The estimate above was obtained within the rigid-texture approximation, in which the magnetic field changes the energy of the skyrmion crystal without significantly modifying its characteristic wave vector. This assumption can be relaxed by allowing the uniform susceptibility of the skyrmion crystal to acquire a weak dependence on the 
skyrmion density $\rho_{\rm sk} \sim Q^2$. 
In the long-wavelength regime, locality and the absence of scalar invariants linear in gradients imply the following expansion for the magnetic susceptibility $\chi_m$ of the skyrmion crystal:

\begin{equation}
\chi_m(\rho_{\rm sk}) = \chi_0 + \chi_1 \rho_{\rm sk} + O(\rho_{\rm sk}^2).
\end{equation}
Including this dependence, the field-dependent energy density becomes

\begin{eqnarray}
e(\rho_{\rm sk},h) &=& \frac{\sqrt{3}}{4} a(\gamma)+\frac{3}{8}b\rho_{\rm sk}+\frac{9\sqrt{3}}{64}c\rho_{\rm sk}^2 
\nonumber \\
&-& \frac{1}{2}\left[\chi_0 + \chi_1 \rho_{\rm sk} \right] h^2.
\end{eqnarray}
The magnetic field therefore renormalizes the coefficient of the linear term,
\begin{equation}
a \;\to\; a_{\rm eff}(h) = a - \frac{2}{\sqrt{3}}\chi_1 h^2,
\end{equation}
so that minimization with respect to $\rho_{\rm sk}$ yields

\begin{equation}
\rho_{\rm sk}^{\ast}(h) = -\frac{2\sqrt{3}}{3}\frac{a_{\rm eff}(h)}{b}
= -\frac{2\sqrt{3}a}{3b} + \frac{4\chi_1}{3b} h^2 + O(h^4),
\label{eq:density-lwl}
\end{equation}
where we used $b>0$ close to $\gamma_m$. Thus, the skyrmion density acquires a quadratic dependence on the applied field
which physically slightly lowers the crystal's energy slightly prior to saturation.
Importantly, 
this field-induced correction does not modify the phase boundary estimated in Eq.~\eqref{eq:sat2} at leading order;
it does not alter the fundamental level-crossing mechanism that 
forces the abrupt phase transition.
Because the saturation field approaches zero near the transition threshold 
($\chi_0 h_{sat} \to 0$ as $\gamma \to \gamma_m$), the skyrmion crystal is destroyed before any meaningful field-induced density shift can occur. Therefore, the rigid-texture assumption becomes exact in this limit. The resulting phase diagram in the long-wavelength limit is shown in 
Figure ~\ref{fig:longwavelength_pd} and compared against unbiased numerical simulations in Fig.~\ref{fig:shortwavelength_pd}.

To summarize, the saturation transition is not controlled by the continuous vanishing of the skyrmion density. Numerically, $Q^2$ changes very weakly with field and then drops discontinuously to zero at $h_{\rm sat}$, while the magnetization increases only modestly up to the transition. This behavior indicates that the double-skyrmion crystal remains remarkably rigid under the applied field and loses stability through a strongly first-order transition into the fully polarized state. The magnetic field does, however, induce a weak deformation of the skyrmion cores that lowers the energy of the skyrmion crystal. As a result, Eq.~\eqref{eq:hsat}, obtained within the rigid-texture approximation, provides a lower bound for the saturation field.

Finally, the zero-field density scaling $\rho_{sk} \sim (\gamma - \gamma_m)$ in Eq.~\eqref{eq:est} implies that the ordering wave vector emerges as $Q \sim \sqrt{\gamma - \gamma_m}$. 
While this critical scaling behavior formally analogous to a that at
a Pokrovsky-Talapov transition, the
underlying structure of the two problems is fundamentally different.
In the Pokrovsky--Talapov scenario, the incommensurate phase is
described as a dilute gas of solitons, characterized by two distinct
length scales: the soliton core size, which remains finite at the
transition, and the average separation between solitons, which diverges
as the transition is approached. 

By contrast, the present instability is governed by a single wavevector
$Q$ that controls both the spatial variation
of the texture and the separation between its features. As $Q \rightarrow 0$ at the transition, the entire structure uniformly dilates,
and there is no regime in which well-separated, localized objects can
be identified. The skyrmion crystal should therefore be viewed as a long-wavelength instability of the fully polarized state, characterized by a vanishing 
modulation wavevector, rather than as a dilute assembly of localized topological particles with a fixed core size.

\section{Full Phase Diagram}
\label{sec:Phase_Diagram}

The long-wavelength analysis of Sec.~\ref{sec:holomorphic_to_crystal} provides a controlled description of the skyrmion crystal in the regime 
$Q \ll  1/a_0$ where
$a_0$ is the lattice constant; here the ordering wave vector 
$Q$ evolves continuously from zero, and the skyrmion density is 
set by the
competition between the topological chemical potential and the lattice gradient corrections. However, as the chiral coupling $\gamma$ increases and $Q \to 1/a_0$, this continuum description breaks down and discrete lattice effects become dominant.

In this section, we determine the global phase diagram of the lattice model
by combining analytical arguments with numerical minimization of the microscopic Hamiltonian. We explicitly demonstrate how the lattice phase diagram connects to the long-wavelength theory of Sec.~\ref{sec:holomorphic_to_crystal}, and we characterize the sequence of commensurate double skyrmion crystal states (Sec.~\ref{sec:dblskx}) and their evolution into a tetra-skyrmion crystal phase (Sec.~\ref{sec:tetraskx}).
 
\subsection{Double Skyrmion Crystals \label{sec:dblskx}} 

\begin{figure}[!t]
    \centering
    \includegraphics[width=0.45\textwidth,keepaspectratio]{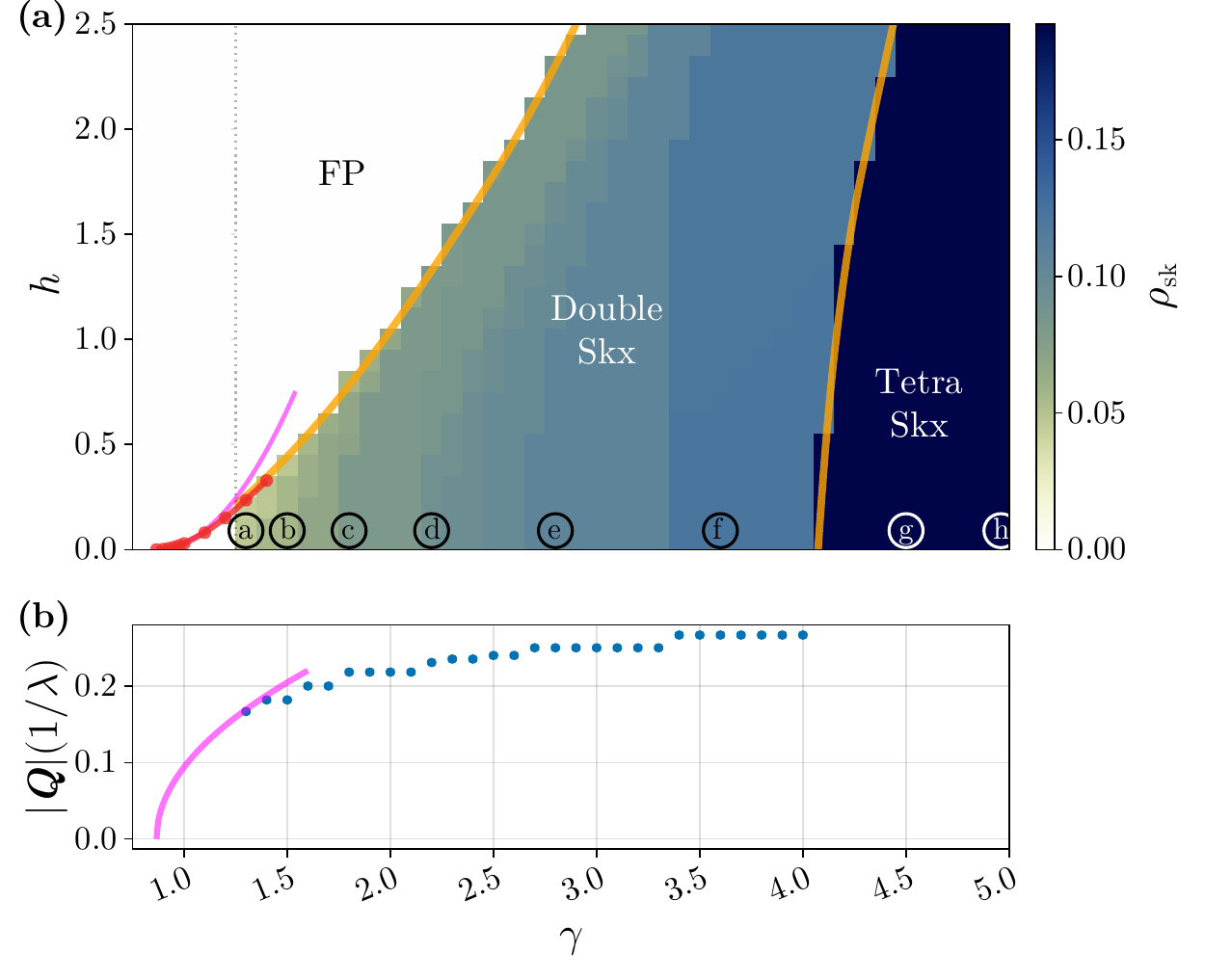}
    \caption{%
    (\textbf a) Phase diagram for the Hamiltonian of Eq.~\eqref{eq:central} as determined from numerical simulations on a lattice. The color gradient, drawn from a grid of unbiased simulations beginning with $\gamma=1.3$, represents the skyrmion density $\rho_{\rm sk}$. With increasing $\gamma$, the phases develop from a fully polarized (FP) order into a double skyrmion crystal (Double Skx). The skyrmion density increases with larger $\gamma$ until finally transitioning into a tetra-skyrmion state with uniform scalar spin chirality and four skyrmions per magnetic unit cell (Tetra Skx). The phase boundary between the FP and Double Skx phases illustrated in Fig~\ref{fig:unbiased_near_gammam} (d) is shown in red dots and lines. The prediction of the long wavelength theory of Eq.~\eqref{eq:hsat}, with parameters fit to the double skyrmion ansatz, is shown in solid magenta; departures from the numerical results begin to appear around $\gamma=1.3$. The circled points (a)--(h) indicate the points of the phase diagram used to calculate the static and dynamic structure factors illustrated in Fig.~\ref{fig:dynamics}. Point (h) is understood to lie at the limit $\gamma \to\infty$ (equivalently, $J=0$), where the Tetra Skx phase becomes strictly degenerate with the tetrahedral phase.
    (\textbf b) 
    The skyrmion lengthscale found numerically in the Double Skx portion of the phase diagram at $h=0$ is shown as blue dots. The predicted $Q(\gamma)$ as determined by the long wavelength theory [Eq.~\eqref{eq:Qgamma}] using parameters fit to the double-skyrmion ansatz is shown in magenta.
    As $Q$ increases, lattice effects become important leading to Frenkel-Kontorova pinning and a series of tunable, commensurate plateaus.
    }
    \label{fig:shortwavelength_pd}
\end{figure}

Determination of the global phase diagram (Fig.~\ref{fig:shortwavelength_pd}(a)),
spanning from the macroscopic continuum ($Q \ll 1/a_0$) down to the discrete lattice scale ($Q \sim 1/a_0$), requires a two-pronged numerical approach.
In the long-wavelength regime near $\gamma_m$, the characteristic wave vector $Q$ of the skyrmion crystal becomes arbitrarily small. Because the corresponding   magnetic unit cell of the real-space texture spans an enormous number of lattice sites, purely unbiased simulations are computationally prohibitive. In this domain ($\gamma  \le 1.5$), the phase boundary was determined by numerically optimizing the analytical double-skyrmion ansatz  (Eq.~\eqref{eq:tripleQansatz}) via the conjugate-gradient algorithm over a grid of integer-valued length scales $\lambda = 1/Q$ and a set of applied fields $h$. Details are given in Appendices \ref{sec:app_coeff_fitting} and \ref{sec:app_num_lw_ansatz}.

\begin{figure}[t]
    \centering
    \includegraphics[width=1.0\columnwidth]{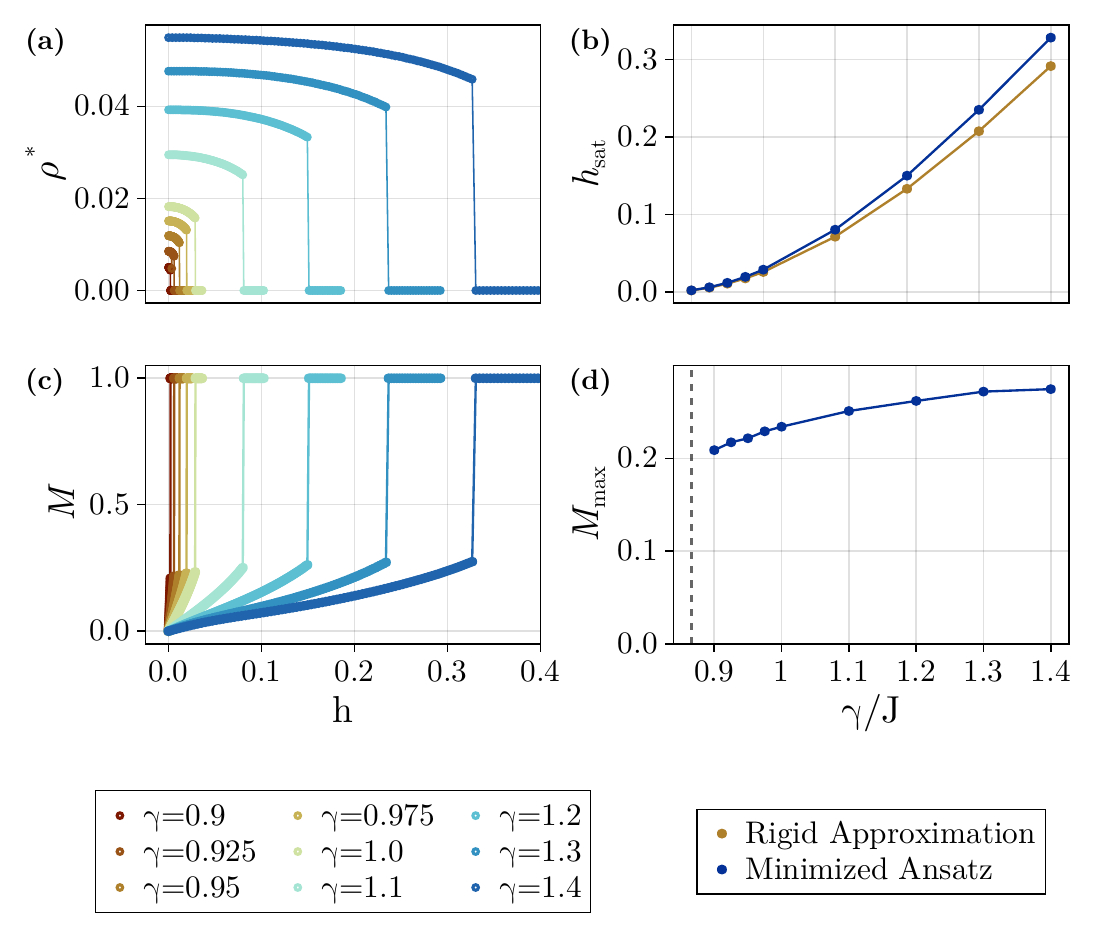}
    \caption{
    Behavior of the optimized double skyrmion ansatz approaching the transition to the fully polarized state. 
    (\textbf a) Evolution of the skyrmion density $\rho_{\rm sk}$ as a function of field for a range of small $\gamma$ values above the transition. 
    (\textbf b) Phase boundary determined by an interpolated numerical study of the ansatz (blue dots) together with the lower-bound on the phase boundary predicted by the rigid approximation [Eq.~\eqref{eq:hsat}]. The rigid approximation remains good up to about $\gamma=1.1$.
    (\textbf c) Evolution of the magnetization $M$ as a function of field for a series of $\gamma$ values.
    (d) Magnetization of the ansatz just below the saturation field $h_{\rm sat}$.}
    \label{fig:unbiased_near_gammam}
\end{figure}

Conversely, as the chiral coupling increases deeper into the ordered phase ($\gamma \ge 1.3$), the characteristic wave vector grows toward the inverse lattice spacing ($Q \rightarrow 1/a_0$). Here, discrete lattice effects become dominant, requiring a fully unbiased numerical strategy to capture the resulting commensurate states. To avoid imposing artificial commensurability constraints on the ordering wave vector, spin configurations were repeatedly initialized on variable-size supercells (with randomly sampled linear dimensions up to $L=50$) using randomized, infinite-temperature states. These were subjected to simulated annealing via stochastic Landau-Lifshitz-Gilbert (LLG) dynamics, followed by zero-temperature gradient optimization. This procedure was repeated many times together with a systematic comparison of candidate spin configurations at each point with re-optimized candidates at neighboring points of the phase diagram, ensuring propagation of the lowest energy solutions to related regions of the phase diagram. Algorithmic details are provided in Appendix ~\ref{sec:app_num_pd}. Crucially, in the crossover region ($1.3 < \gamma < 1.5$) where these two distinct numerical frameworks overlap, they yield perfectly consistent phase boundaries.

Consistent with the long-wavelength analysis, the double-skyrmion phase is stabilized over a broad parameter window, $\gamma_m<\gamma \leq 4$. However, as the chiral coupling increases and the skyrmion density reaches $\rho_{\rm sk}\simeq 0.01$ (corresponding to $Q\sim 0.1/a_0$), the quasi-continuous evolution of the ordering wave vector fundamentally changes. In this lattice-dominated regime, the ordered state can no longer be viewed as a smooth deformation of the continuum skyrmion crystal. As shown explicitly in Fig.~\ref{fig:shortwavelength_pd}(b), $Q$ locks into discrete commensurate plateaus as a function of $\gamma$, whose widths grow as the ordering wave vector approaches the inverse lattice spacing. This staircase-like evolution is the magnetic analogue of Frenkel--Kontorova physics~\cite{KontorovaFrenkel1938,Aubry1983,Bak1982}: the competition between the preferred skyrmion density selected by the topological chemical potential and pinning by the underlying discrete lattice drives the system through a sequence of commensurate states.

This robust lattice pinning fundamentally restricts the system's response to an applied magnetic field. As visually highlighted by the vertical color bands in
Fig.~\ref{fig:shortwavelength_pd}(a) and explicitly plotted in 
in Fig.~\ref{fig:unbiased_near_gammam}(a), this rigidity is especially pronounced at low skyrmion densities ($\rho_{sk} \simeq 0.01$), where the density remains remarkably constant before dropping discontinuously to zero at the saturation field ($h_{sat}$). This abrupt transition to the fully polarized state is captured by the numerically determined phase boundary in Fig.~\ref{fig:unbiased_near_gammam} (b), which closely tracks the analytical lower bound predicted by the rigid approximation of 
Eq.~\eqref{eq:sat2}. 
The pronounced rigidity of the skyrmion crystal is also reflected in its 
magnetization (Figs.~\ref{fig:unbiased_near_gammam}(c) and (d)) which weakly increases to only $\sim 20\% $ of its saturation value just prior to $h_{sat}$. Rather than altering the topological density, the applied field simply forces a local contraction of the spins pointing antiparallel to the field. Thus, 
$\gamma$ tunes the preferred skyrmion density, while $h$ primarily deforms the texture at nearly fixed density until the first-order saturation transition.

\subsection{The Lattice Regime:  Tetra-Skyrmion and Tetrahedral Orders \label{sec:tetraskx}} 

As the wavelength approaches the discrete lattice scale, the resulting magnetic states are best analyzed in real space.
For larger $\gamma$, the optimal wavelength decreases further and the skyrmion cores begin to strongly overlap, signaling the breakdown of the long-wavelength description. As the continuous winding of the topological charge is inevitably discretized, the semiclassical framework gives way to a regime governed entirely by local lattice geometries.

This motivates examining magnetic states that emerge directly at the lattice scale when the chiral interaction strongly dominates ($\gamma \gg J$). In this extreme short-wavelength limit, the zero-field magnetic orders are fundamentally dictated by the requirement to maximize the local scalar spin chirality on every plaquette. It is important to emphasize that in the limit  $\gamma  \to \infty$, all spin configurations that achieve this uniform, maximal scalar chirality become exactly degenerate. Consequently, it is the subdominant Heisenberg interaction that must ultimately act as a selection mechanism, lifting this extensive degeneracy to stabilize a specific ordering pattern.

\begin{figure*}[t]
    \centering
    \includegraphics[width=0.9\textwidth]{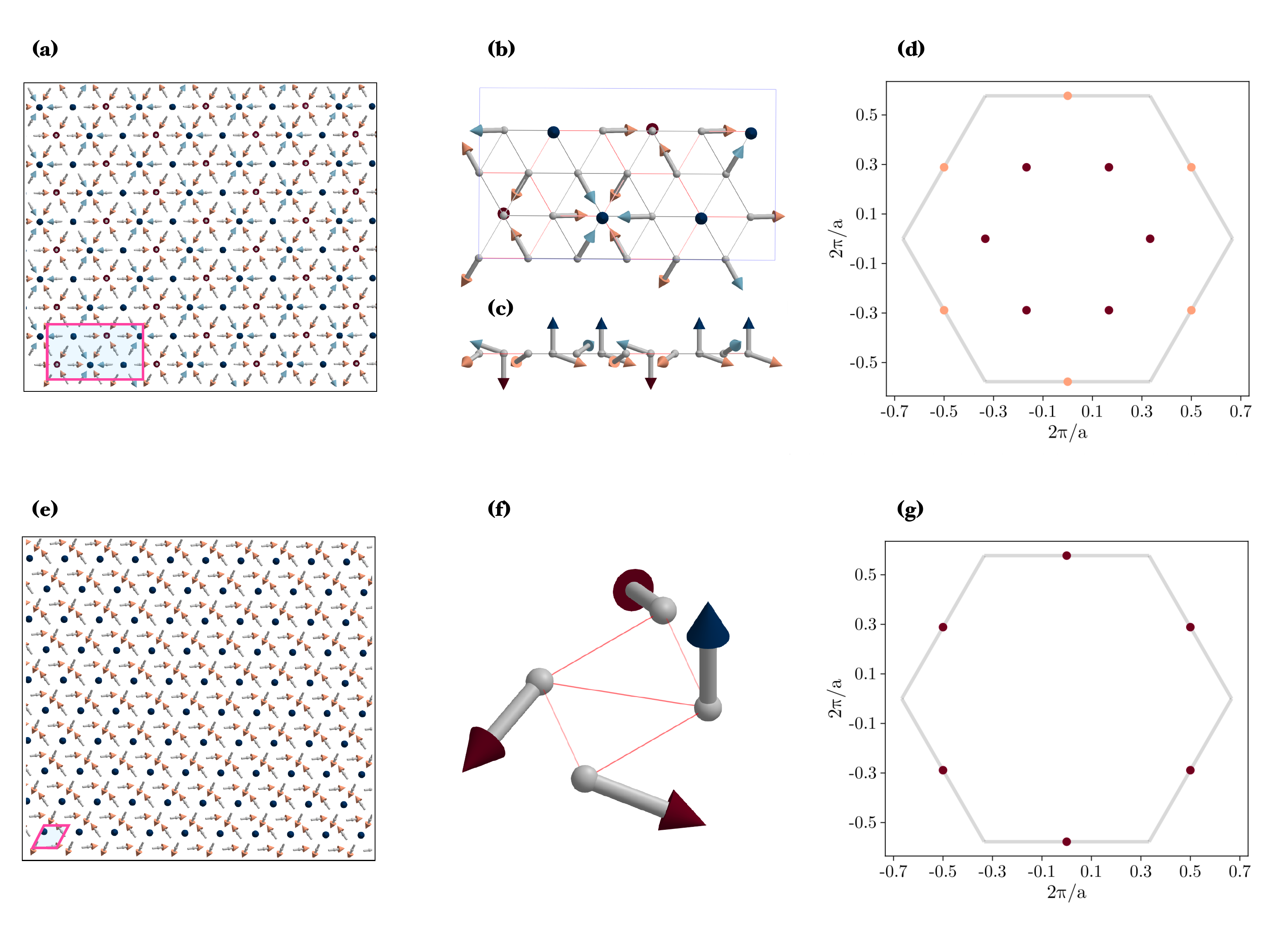} 
    \caption{\textbf{Tetra-Skyrmion Crystal (a--d)}: As $\gamma$ increases, the phase diagram converges to a tetra-skrymion order, a 24-spin structure with a topological charge of four, illustrated in (a). A single magnetic cell of the texture (inset of (a)) is shown in othographic projection from above (b) and the side (c).  The value of the inner product, $S_i\cdot S_j$ on each bond is either $+1/3$ (gray) or $-1/3$ (orange).
    (d) The static structure factor, $\sum_{\alpha}\mathcal{S}^{\alpha\alpha}\left({\bm q}\right)$, is characterized by three wave vectors with significant weight at the midpoints between the $\Gamma$ and K points; additional weight is found at a secondary set of wave vectors located at the M points. The combined weights of the midpoint wave vectors are $24N_sS^2/27$, while those at the M points are $3N_sS^2/27$, where $N_s$ is the number of sites. No higher harmonics are present. \textbf{Tetrahedral Order (e--g): }In the limit $h\to0$ and $\gamma \to\infty$ limit, the tetra-skrymion phase becomes strictly degenerate with the tetrahedral order, shown here. Both solutions yield exactly the same scalar spin chirality $\chi_{ijk}=S_i\cdot (S_j\times S_k$) on each plaquette. The tetrahedral spin texture is shown in (e), where a single magnetic cell is indicated in magenta. The four-spin magnetic cell is illustrated in (f). The static structure factor, $\sum_\alpha \mathcal{S}^{\alpha\alpha}\left({\bm q}\right)$, is characterized by weight equally distributed on the M points, illustrated in (g).}
    \label{fig:tetraskx_tetrahedral}
\end{figure*}

Our unbiased numerical simulations indicate that the selected state is a previously unidentified tetra-skyrmion crystal phase whose characteristic length scale is only a few lattice spacings. As illustrated in Fig.~\ref{fig:tetraskx_tetrahedral} (a), (b) and (c), the magnetic unit cell of this highly novel topological order contains 24 spins, corresponding to 48 elementary triangular plaquettes. 

In order to connect this discrete lattice geometry to the macroscopic topological charge, we recall that the continuum skyrmion number, $N_{\rm sk} = \frac{1}{4\pi} \int \mathbf{n} \cdot (\partial_x \mathbf{n} \times \partial_y \mathbf{n}) d^2r$, translates on the lattice to the normalized sum of the oriented solid angles subtended by the spins on each elementary triangle. A striking feature of the tetra-skyrmion phase is that the three spins on every single triangle subtend the exact same solid angle,
$\Omega_0 = \frac{\pi}{3}.$
Geometrically, this means the local spin texture of each plaquette maps out exactly one-twelfth of the target order-parameter sphere ($4\pi$), thereby producing a strictly uniform topological charge density. Summing this uniform coverage over the 48 plaquettes of the magnetic unit cell yields a total wrapped solid angle of $16\pi$. The total skyrmion number per unit cell is therefore an unusual
\begin{equation}
    N^{\rm uc}_{\rm SkX} = \frac{48\,\Omega_0}{4\pi} = 4.
    \end{equation}
The emergence of such a high topological charge within a single zero-field unit cell underscores the distinct physics of this lattice-dominated regime, setting it sharply apart from the conventional $N=1$ or $N=2$ continuum skyrmion crystals.

The Heisenberg energy reflects this remarkable geometric rigidity. The nearest-neighbor correlations take only two values, as indicated by the color of the bonds in Fig.~\ref{fig:tetraskx_tetrahedral} (b): $\mathbf{n}_i\cdot\mathbf{n}_j=1/3$ on the grey bonds and $\mathbf{n}_i\cdot\mathbf{n}_j=-1/3$ on the orange bonds. Since each triangle contains two grey bonds and one orange bond, the Heisenberg energy per site is $e_H = -\frac{J}{3}$.
    Because the scalar chirality is uniform, the chiral contribution is
    $e_\chi = -2\gamma \chi_t = -\frac{8\gamma}{3\sqrt{3}}$.
    The total energy per site of the zero-field tetra-skyrmion crystal is therefore
    \begin{equation}
        e = -\frac{J}{3} - \frac{8\gamma}{3\sqrt{3}}.
        \end{equation}
This low energy density demonstrates how the 24-spin lattice geometry accommodates the topological demands of the chiral interaction while minimizing frustration in the Heisenberg exchange.  Because this energy is strictly locked to the rigid real-space geometry of the 24-spin unit cell, the resulting magnetic texture leaves a highly distinct, commensurate signature in momentum space.

The static spin structure factor, shown in Fig.~\ref{fig:tetraskx_tetrahedral} (d), exhibits dominant weight at six symmetry-related wave vectors located midway between the $\Gamma$ point and the corners (K points) of the hexagonal Brillouin zone, together with subdominant contributions at the M points corresponding to the midpoints of the zone edges. Because of this simple structure, the phase admits an analytical expression. The full real-space parametrization of this 24-site magnetic unit cell and the resulting tetra-skyrmion state is discussed in detail in  Appendix~\ref{sec:app_tretra-SkX}.

\allowdisplaybreaks

In the asymptotic limit where the chiral interaction entirely dictates the energy ($\gamma \rightarrow \infty$ or $J \rightarrow 0$), the tetra-skyrmion crystal and the conventional four-sublattice tetrahedral state become exactly degenerate. This degeneracy arises because both configurations globally maximize the local scalar spin chirality on every elementary triangular plaquette, thereby achieving the exact same, absolute minimum of the chiral energy. The conventional four-sublattice tetrahedral state is contained in the variational family defined by Eqs.~\eqref{eq:tripleQansatz} and \eqref{eq:Mpoints} as its shortest-wavelength limit. Indeed, for $\lambda=2$, the three ordering wave vectors become $\mathbf Q_\nu=\mathbf M_\nu$, coinciding with the three symmetry-related $M$ points of the Brillouin zone. The cosine components in Eq.~\eqref{eq:tripleQansatz} then take the values $\pm1$ on the lattice sites, producing four spin orientations directed toward the vertices of a regular tetrahedron. The eight elementary triangles in the corresponding magnetic unit cell each subtend the same oriented solid angle $\Omega=\pi$, yielding a total topological charge $|N_{\rm sk}|=8\pi/(4\pi)=2$, consistent with its identification as the shortest-wavelength limit of the double-skyrmion crystal.

While the ferromagnetic Heisenberg interaction ($J>0$) selects the 24-spin
tetra-skyrmion crystal, antiferromagnetic exchange ($J<0$) lifts the
asymptotic degeneracy in favor of tetrahedral order. More generally, this
order is selected whenever the Fourier transform $J(\mathbf{q})$ has global
minima at the three $M$ points, $\mathbf{q}=\mathbf{M}_\nu$ with
$\nu=1,2,3$~\cite{Martin2008,Batista2016}. Remarkably, this classical result
has a quantum counterpart: density matrix renormalization group calculations
for an $S=1/2$ version of the triangular-lattice Hamiltonian with
antiferromagnetic exchange and the same scalar-chirality interaction identify
the same ground-state order~\cite{Gong2017}. This agreement is nontrivial
because no small parameter controls the classical approximation at $S=1/2$
and short wavelengths, demonstrating that tetrahedral order survives deep
in the quantum regime.

Tetrahedral order also arises in itinerant triangular Kondo-lattice models,
at $3/4$ filling in the weak-coupling limit~\cite{Martin2008} and at $1/4$
filling in the intermediate-coupling regime~\cite{Akagi2010}. These results
connect the short-wavelength tetrahedral phase of our model to the
spontaneous triple-$Q$ order recently observed in metallic triangular
antiferromagnets such as
$\text{Co}_{1/3}\text{TaS}_2$~\cite{Takagi2023,Park2023}.
\section{Dynamical Structure factors}
\label{sec:experiment}

\begin{figure*}
    \centering
\includegraphics[width=0.9\textwidth]{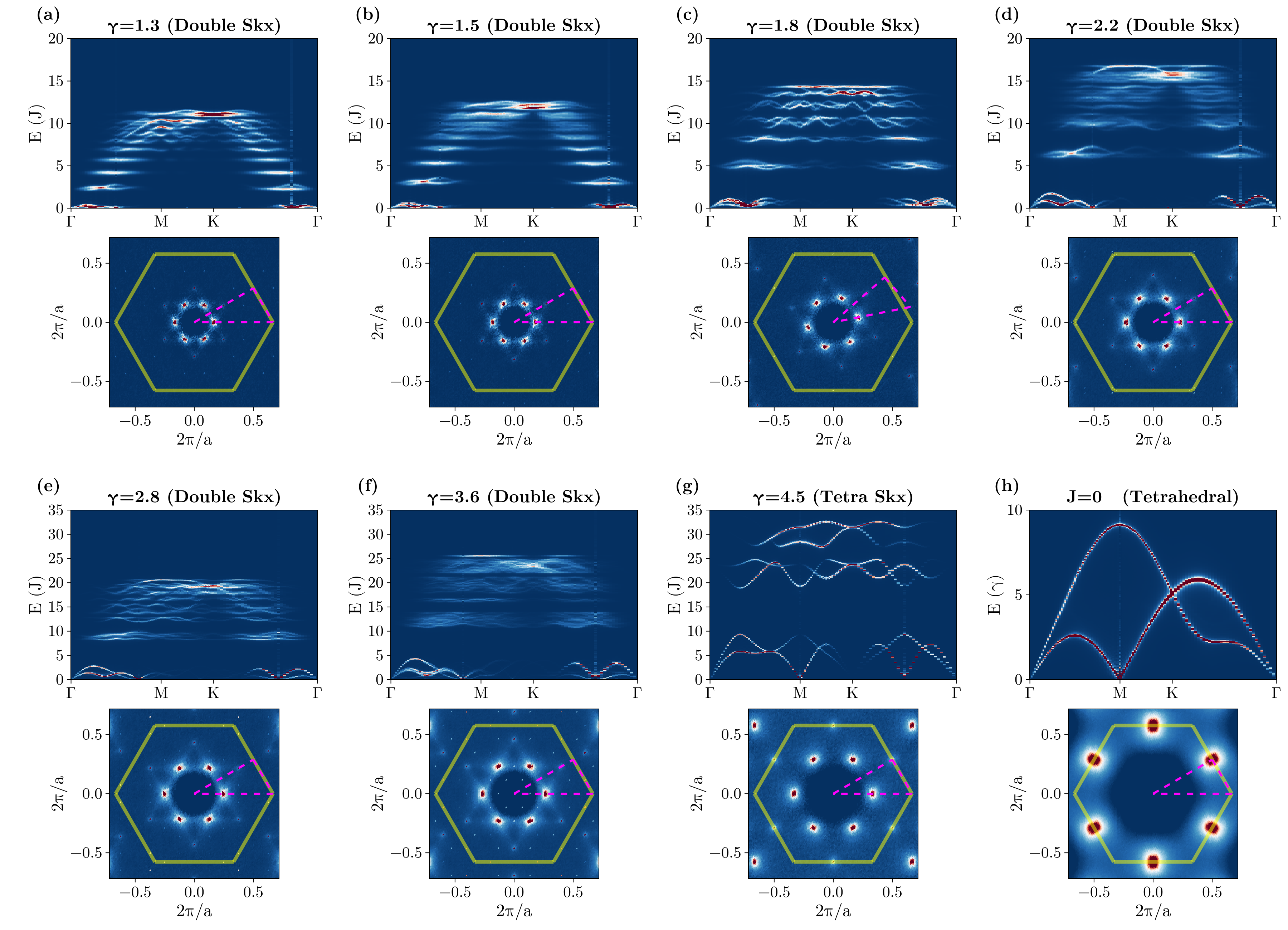}
    \caption{
    Representative dynamical (top of labeled pairs) and static (bottom of labeled pairs) spin structure factors taken along the $h=0$ axis of the phase diagram shown in Fig.~\ref{fig:shortwavelength_pd}. The labels here correspond to the labels of that figure. Calculations were performed at a temperature of $T=0.03J$ using stochastic Landau-Lifshitz-Gilbert dynamics, starting from ground states determined using the unbiased approach described in Appendix~\ref{sec:app_num_pd}. Details are given in Appendix~\ref{app:DSSF}. The Double Skyrmion phase (Sec.~\ref{sec:dblskx}) is illustrated in (a) through (f); the dilation of the dominant wave vector along the $\Gamma-M$ line with increasing $\gamma$ is clearly visible. 
    This region is characterized by Frenkel-Kontorova physics~\cite{KontorovaFrenkel1938}, where the topological skyrmion potential competes with discrete lattice effects. One characteristic feature in this regime is the occasional decoupling of the reciprocal lattice vectors of the skyrmion crystal from the reciprocal lattice vectors of the microscopic lattice, made evident in panel (c) as a global rotation of the dominant wave vectors. Note that the high-symmetry path used to generate the dynamical structure factor is rotated to align with the dominant wave vectors in this case. The final two panels show the very short-wavelength phases, namely (g) the tetra-skyrmion phase, and (h) the tetrahedral phase (both illustrated and described in Fig.~\ref{fig:tetraskx_tetrahedral} and Appendix~\ref{sec:tetraskx}). 
    }
    \label{fig:dynamics}
\end{figure*}

 The nature of the phase transitions driven by the chiral coupling is revealed by the evolution of the ordering wave vectors. Direct identification of the resulting skyrmion phases, however, poses a fundamental challenge for elastic neutron diffraction. In particular, diffraction cannot generally distinguish an intrinsic coherent triple-$\mathbf Q$ state from a multidomain single-$\mathbf Q$ state: after averaging over the three symmetry-related domains, the latter produces Bragg peaks at the same ordering wave vectors as the former. Consequently, the observation of a triple-$\mathbf Q$ diffraction pattern does not by itself establish the intrinsic multiple-$\mathbf Q$ character, noncoplanarity, or topology of the underlying spin configuration.

This ambiguity can be resolved through dynamical measurements, as
recently emphasized for multiple-$\mathbf Q$ orderings in frustrated
magnets~\cite{Park2025}. A single-$\mathbf Q$ spiral retains a
generalized translation symmetry: translation by a lattice vector
$\mathbf R$ can be compensated by a spin rotation through an angle
$\mathbf Q\cdot\mathbf R$ about the spiral axis. Its spin-wave spectrum
can therefore be unfolded into the crystallographic Brillouin zone,
with a single fundamental magnon dispersion contributing to the 
dynamical spin structure factor
(DSSF)
at $\mathbf q$ and $\mathbf q\pm\mathbf Q$, thereby producing only
three magnon branches. A multidomain single-$\mathbf Q$ state produces
an incoherent superposition of the corresponding spectra from the
symmetry-related domains. By contrast, an intrinsic triple-$\mathbf Q$
state possesses no analogous combined translation--spin-rotation
symmetry. Its enlarged magnetic unit cell and the coupling among its three
order-parameter components produce extensive mode folding and
hybridization, resulting in the multiple magnon bands illustrated in
Fig.~\ref{fig:dynamics}(a)--(f). These spectra were computed using
low-temperature Landau--Lifshitz dynamics and converted into the
corresponding quantum spin-wave response using the standard
quantum--classical correspondence factor; details are provided in
Appendix~\ref{app:DSSF}. The resulting DSSF provides experimentally
accessible fingerprints for distinguishing these skyrmion phases
using inelastic neutron scattering and related momentum-resolved
probes~\cite{wei2018electrical,assouline2021excitonic}.

Besides the multiple magnon minibands already mentioned, which arise from the large magnetic unit cell of the double-SkX, Fig.~\ref{fig:dynamics} reveals several general features. All three phases spontaneously break the continuous $SO(3)$ spin-rotation symmetry completely and therefore exhibit three Goldstone modes associated with long-wavelength rotations of the noncollinear order parameter. Beyond this universal low-energy structure, the different multiple-$\mathbf Q$ textures produce qualitatively distinct distributions of spectral weight throughout the Brillouin zone, reflecting their different magnetic unit cells, ordering wave vectors, and topological spin structures.

Importantly, these calculations provide a direct illustration of the variation in wavevector $Q$. As we tune $\gamma$ (Fig.~\ref{fig:dynamics} (a)--(h)), the wavelength seen by the $\omega=0$ resonance at finite momentum shifts towards the $M$ points. Here we observe a defining change in the collective modes between the tetra-skyrmion crystal ordering (Fig.~\ref{fig:dynamics} (g)) and the more familiar tetrahedral order (Fig.~\ref{fig:dynamics} (h)). Whereas in the tetrahedral order the finite-momentum folding (and instability) occurs only and strictly at the $M$ point, the tetra-skyrmion has an additional, strongly weighted $\omega=0$ state at a finite momentum on the $\Gamma-K$ line.

The spectrum of magnons also sharply differs between these states. Whereas excitations and vibrational modes are abundant in the double-skyrmion DSSF (characteristic of the large unit cell required to create it), only two principal magnon branches are found in the tetrahedral order. The presence of nested excitations would be a decisive signature of strong folding, which we anticipated in the double-skyrmion crystal phase. These rich spectral features demonstrate 
that the real-space topology of these skyrmion phases leaves a definitive, measurable imprint on their momentum-space dynamics. By characterizing these distinct resonant shifts and  magnon-bands structures, our results provide a concrete experimental roadmap towards identifying the microscopic nature of the symmetry-broken cascades observed in fractional Chern insulators.

\section{Discussion}
\label{sec:discussion}

Here we have demonstrated that a minimal model of spin excitations within a Wigner crystal supports tunable zero-field skyrmion crystals, provided the underlying electronic bands possess non-trivial topology.
Specifically, this emerges from an intrinsic scalar-chirality interaction predicated on time-reversal symmetry breaking, which is itself generated by electronic topology.

In the continuum limit, we explicitly demonstrated that the scalar-chirality contribution is akin to a chemical potential fixing the skyrmion number,
$ H_\chi = -8\pi\gamma N_{\rm sk}$.
While such a term shows that an instability towards a skyrmion ground state is possible, lattice scale corrections that break scale-invariance are needed to select the equilibrium wavelength.

This mechanism is qualitatively distinct from the more familiar routes to magnetic skyrmion crystals.
Unlike the charged skyrmions of the quantum Hall ferromagnet, which are stabilized by Coulomb interactions in the presence of a strong external magnetic field \cite{Sondhi93}, the long-wavelength skyrmion crystal here emerges at zero field.
In DMI systems, for example, the skyrmion length scale is set directly by the ratio of exchange to spin-orbit-induced terms, and again a magnetic field is typically required to stabilize the skyrmion crystal over competing helical or conical phases
\cite{nagaosa2013topological,back20202020,Moriya60,
dzyaloshinsky1958thermodynamic}.  Similarly in itinerant RKKY mechanisms, multiple ordering wave vectors are often selected by Fermi-surface structure, so the skyrmion period is tied to microscopic electronic scales
\cite{Ozawa2017,Wang2020,Wang2023}.  
By contrast, reminiscent of the mechanism proposed for chiral Stoner magnetism in Dirac bands~\cite{Dong2024}, the long-wavelength skyrmion crystal in the topological ferromagnetic Wigner crystal emerges through an instability of the fully polarized state. Here, the instability is driven by the Berry-phase-induced scalar-chirality interaction, whose strength relative to the ferromagnetic exchange is encoded in the ratio $\gamma$. Near the
transition, the ordering wave vector evolves continuously, giving a skyrmion density that is tunable rather than fixed by a pre-existing microscopic length scale.

The nature of this instability is also different from a dilute-soliton transition.  Although the scaling of the ordering wave vector near $\gamma=\sqrt{3}/2$ resembles the square-root onset familiar from commensurate-incommensurate problems, the present skyrmion crystal is
controlled by a wavevector $Q$.  This scale determines both the skyrmion core size and density.  Thus, as the transition is approached from the skyrmion-crystal side, the entire texture dilates uniformly rather than decomposing into well-separated solitons with fixed core size.  We note that this single-scale structure allows the continuum energy density to be organized as an expansion in powers of $Q$ or, equivalently, in powers of $\rho_{\rm sk}$.

At larger $\gamma$, the continuum description crosses over into a lattice-controlled regime.  The smooth evolution of the skyrmion density is replaced by commensurate plateaus, reflecting the locking of the magnetic unit cell to the underlying triangular lattice.  At the end of this sequence, at 
$\gamma \approx 4$, is the tetra-skyrmion crystal, a 24-site magnetic unit cell carrying total topological charge $|N_{\rm sk}|= 4$.  This state has uniform scalar spin chirality on every elementary plaquette and becomes degenerate, in the limit $\gamma\to\infty$, with the four-sublattice tetrahedral order that occurs naturally for antiferromagnetic exchange \cite{Gong2017}.  The result provides a useful bridge between long-wavelength skyrmion crystals and lattice-scale noncoplanar orders previously studied in triangular itinerant magnets \cite{Martin2008,Akagi2010,Park2023,Takagi2023}.

It is useful to distinguish the universal aspects of this mechanism from those that depend on microscopic details. Within the broad class of models describing competition between ferromagnetic exchange and a scalar-chirality interaction, the long-wavelength instability is governed by the generic expansion in the skyrmion density $\rho_{\rm sk}$. Consequently, increasing $\gamma$ produces either a continuous transition for $b>0$, with $\rho_{\rm sk}$ growing continuously from zero, or a first-order transition for $b<0$, with $\rho_{\rm sk}$ jumping to a finite value before increasing further. As lattice discreteness becomes relevant, the competition between the preferred skyrmion-crystal period and the microscopic lattice is also generically expected to produce Frenkel--Kontorova locking, manifested as commensurate plateaus in the ordering wave vector and skyrmion density. These qualitative features are independent of the particular lattice considered. By contrast, the critical coupling, the detailed evolution and plateau structure of $\rho_{\rm sk}$, and especially the phases reached in the short-wavelength regime at large $\gamma$ are nonuniversal and depend on the lattice geometry and microscopic interactions. For example, in the present triangular-lattice model, the tetra-skyrmion crystal and tetrahedral order emerge because they maximize the scalar chirality on the elementary triangles. Different lattice geometries impose different local constraints and will therefore generally select different phases in the short-wavelength limit.

Several signatures follow directly from our calculated phase diagram.  First, the
ordering wave vector of the double-skyrmion crystal should evolve
continuously near the onset transition and then lock into commensurate
plateaus at shorter wavelengths.  Second, an out-of-plane magnetic field
does not primarily tune the skyrmion density in the long-wavelength
regime; instead, it weakly distorts the spin texture until the skyrmion
crystal undergoes a first-order transition into the fully polarized
state.  Third, the different ordered phases have distinct dynamical spin
structure factors, as shown in Fig.~\ref{fig:dynamics}.  These spectra
provide a practical way to distinguish the double-skyrmion,
tetra-skyrmion, and tetrahedral states, particularly because their
elastic structure factors can share similar multiple-$Q$ features.  This
makes dynamical probes such as inelastic neutron scattering, spin
resonance, and related momentum-resolved probes especially valuable for
identifying the underlying magnetic texture
\cite{Park2025,Dahlbom2025}.

Transport provides a complementary diagnostic, although its
interpretation in candidate Chern materials may be subtle.  A noncoplanar
spin texture produces an emergent real-space gauge flux for itinerant
electrons and can therefore contribute to a topological Hall response
\cite{Martin2008,Akagi2010,Ohgushi00}.  In the present setting, however,
the same electronic Berry curvature that helps generate the chiral spin
interaction may also produce an intrinsic anomalous Hall response.  An
important experimental challenge will therefore be to separate the
contribution associated with the pre-existing electronic Berry curvature
from the additional response generated by the emergent skyrmion texture.
The predicted tunability of the ordering wave vector with $\gamma$
offers one possible diagnostic: a Hall contribution that tracks the
skyrmion density would provide strong evidence for the mechanism
proposed here.

Quantum and thermal fluctuations provide a natural direction for
future studies building on our results.  The
classical approximation is best justified for large local moments or for spin textures involving many microscopic spins.  Near the long-wavelength onset transition this description should be especially controlled because the ordering wavevector is small.  In the dense tetra-skyrmion and tetrahedral regimes, however, quantum fluctuations may be more important.  The persistence of related tetrahedral order in spin-$1/2$ triangular-lattice models suggests that at least some of the short-wavelength noncoplanar physics can survive deep into the quantum
regime \cite{Gong2017}.  It would be useful to study whether the continuously tunable double-skyrmion crystal has a quantum counterpart, and whether quantum fluctuations can stabilize adjacent chiral spin liquid or fractionalized phases.

As for the tunability of the Berry flux, we point out that $N\pi$ Berry phases in valley/spin polarized states are a hallmark feature of rhombohedral graphene \cite{Koshino2009,han2024large}, moir\'e superlattices \cite{perea2025quantum,Wang2025} as well as rhombohedral-twisted van der Waals materials; here $N$ is the number of layers. It is further possible that certain rare-earth compounds may host non-trivial magnetic textures \cite{kurumaji2025electronic,gazzah2026near,huang2024hidden}.
The specific proposal here is that tuning the topological content of the electronic bands in the Wigner crystal state can be accomplished by varying $N$ for example, in rhombohedral graphene. Our prediction is that the ground state of the system with a large $N\pi$ Berry phase will destabilize the uniform ferromagnet that arises through valley polarization \cite{Dong2024,parra2025band,zhou2021half}. Detailed calculations of the ferromagnetic exchange $J$ and the three-particle $\gamma$ of such systems is left for future work. 

In conclusion, our results identify the Berry curvature as the active driver of topological magnetic order. Rather than viewing skyrmions only as sources of emergent electronic gauge fields, the present work emphasizes the reciprocal process whereby electronic topology can
generate the chiral spin interactions that stabilize and modulate skyrmion crystals.  This perspective suggests a route to electrically tunable skyrmion matter in correlated two-dimensional materials, where
band topology, symmetry breaking, and magnetism are intrinsically intertwined.

\begin{acknowledgments}
We thank Kipton Barros, Benoit Doucot, Filippo Gaggioli, Daniele Guerci, Sundeep Joy and Leonid Levitov for fruitful discussions. C.D.B. was supported by the U.S. Department of Energy, Office of Science, Office of Basic Energy Sciences, under Award Number DE-SC0022311. D.D. was supported by the Scientific User Facilities Division, Office of Basic Energy Sciences, U.S. Department of Energy, under contract no. DE-AC0500OR22725 with UT Battelle, LLC. 
DK was supported by an Abrahams Postdoctoral Fellowship of the Center for Materials Theory, Rutgers University and a Zuckerman STEM Fellowship.
This work was performed in part at the Aspen Center for Physics, which is supported by National Science Foundation grant PHY-2210452.
\end{acknowledgments}

\bibliography{References.bib}

\appendix

\section{Effective Chiral Interactions in Kondo and Hubbard Models}
\label{app:kondohubbard}

In this section we derive the scalar-chirality term that appears in
Eq.~\eqref{eq:central} from two microscopic starting points distinct
from the case of Wigner crystals discussed in the main text.  The common structure
is that a spin-independent electronic Hamiltonian with nontrivial
orbital phases generates, after the electronic degrees of freedom are
integrated out, an effective spin Hamiltonian of the form
\begin{equation}
    H_{\rm spin}
    =
    -\sum_{\langle ij\rangle}
    J_{ij}\,\mathbf S_i\cdot \mathbf S_j
    -
    \sum_{\triangle_{ijk}}
    \gamma_{ijk}\,
    \mathbf S_i\cdot
    \left(\mathbf S_j\times \mathbf S_k\right).
    \label{eq:spin_model_general}
\end{equation}
Here $\mathbf S_i$ denotes a classical unit vector, and
$\sum_{\triangle_{ijk}}$ runs over elementary triangular plaquettes with
a fixed orientation, which we take to be counterclockwise.  The sign of
$\gamma_{ijk}$ is therefore tied to this orientation convention.  For a
translation-invariant triangular lattice, $J_{ij}\to J>0$ and
$\gamma_{ijk}\to\gamma$, reducing Eq.~\eqref{eq:spin_model_general} to
the spin model quoted in Eq.~\eqref{eq:central}.

The key point is that the scalar spin chirality
\begin{equation}
    \chi_{ijk}
    \equiv
    \mathbf S_i\cdot
    \left(\mathbf S_j\times \mathbf S_k\right)
    \label{eq:scalar_chirality_def}
\end{equation}
is odd under time reversal and odd under the reversal of the plaquette
orientation.  Consequently, its coefficient must also be odd under time
reversal.  Microscopically, this coefficient is supplied by the
imaginary part of an electronic loop amplitude around the triangle.
Equivalently, it is controlled by the gauge-invariant phase accumulated
by an electron encircling an elementary plaquette.

\subsection{Kondo Lattice}

We first consider a Kondo-lattice description of conduction electrons
coupled to localized moments on the triangular lattice.  To avoid
confusing the microscopic Kondo coupling with the ferromagnetic
Heisenberg coupling $J$ in Eq.~\eqref{eq:central}, we denote the former by $J_K$.  We
take
\begin{equation}
    H_{\rm KL}
    =
    H_0
    -
    J_K
    \sum_i
    \mathbf S_i\cdot \boldsymbol{\sigma}_i ,
    \qquad
    J_K>0 ,
    \label{eq:kondo_lattice_hamiltonian}
\end{equation}
where
\begin{equation}
    H_0
    =
    -\sum_{\langle ij\rangle,\alpha}
    \left(
        t_{ij}\,
        c^\dagger_{i\alpha}c_{j\alpha}
        +{\rm h.c.}
    \right)
    -\mu\sum_i n_i .
    \label{eq:electronic_h0}
\end{equation}
The electron spin density is
\begin{equation}
    \boldsymbol{\sigma}_i
    =
    c^\dagger_{i\alpha}
    \boldsymbol{\sigma}_{\alpha\beta}
    c_{i\beta},
    \label{eq:electron_spin_density_sigma}
\end{equation}
with $\boldsymbol{\sigma}$ the vector of Pauli matrices.  The hopping is
spin independent, so the electronic problem remains $SU(2)$ invariant
in spin space.  Orbital time-reversal symmetry breaking is encoded in
complex hoppings,
\begin{equation}
    t_{ij}=|t_{ij}|e^{i\phi_{ij}},
    \qquad
    t_{ji}=t_{ij}^* .
    \label{eq:complex_hoppings}
\end{equation}
The gauge-invariant phase through an oriented triangle $(ijk)$ is
\begin{equation}
    \Phi_{ijk}
    =
    \arg\!\left(t_{ij}t_{jk}t_{ki}\right)
    =
    \phi_{ij}+\phi_{jk}+\phi_{ki}
    \quad
    \bmod 2\pi .
    \label{eq:plaquette_flux}
\end{equation}

For a static spin configuration, integrating out the electrons gives
\begin{equation}
    S_{\rm eff}[\mathbf S]
    =
    -{\rm Tr}\,
    \ln\!\left[
        G_0^{-1}
        -
        J_K \hat V
    \right],
    \qquad
    \hat V_{ij}
    =
    \delta_{ij}\,
    \mathbf S_i\cdot \boldsymbol{\sigma},
    \label{eq:kondo_effective_action}
\end{equation}
where $G_0=(i\omega_n-H_0)^{-1}$ is the unperturbed electronic Green's
function.  The trace in Eq.~\eqref{eq:kondo_effective_action} is over
lattice sites, Matsubara frequencies, and spin indices.  Dropping a
spin-independent constant,
\begin{equation}
    S_{\rm eff}[\mathbf S]
    =
    \sum_{n=1}^{\infty}
    \frac{J_K^n}{n}\,
    {\rm Tr}\!\left[
        \left(G_0 \hat V\right)^n
    \right].
    \label{eq:kondo_log_expansion}
\end{equation}
The term linear in $\mathbf S_i$ vanishes in the absence of an
electronic spin polarization.  The leading nontrivial contribution is
quadratic:
\begin{align}
    H^{(2)}_{\rm eff}
    &=
    J_K^2
    \sum_{ij}
    \Pi_{ij}\,
    \mathbf S_i\cdot \mathbf S_j ,
    \label{eq:kondo_quadratic_realspace}
    \\
    \Pi_{ij}
    &=
    T\sum_{\omega_n}
    G_0(i,j;i\omega_n)\,
    G_0(j,i;i\omega_n).
    \label{eq:kondo_bubble_realspace}
\end{align}
Equivalently, in momentum space,
\begin{align}
   \notag H^{(2)}_{\rm eff}
    &=
    J_K^2
    \sum_{\mathbf q}
    \Pi(\mathbf q)\,
    \mathbf S_{\mathbf q}\cdot \mathbf S_{-\mathbf q},
    \\
    \Pi(\mathbf q)
    &=
    T\sum_{\mathbf k,\omega_n}
    G_0(\mathbf k,i\omega_n)
    G_0(\mathbf k+\mathbf q,i\omega_n).
    \label{eq:kondo_rkky_momentum}
\end{align}
This is the usual RKKY interaction.  With the sign convention of
Eq.~\eqref{eq:spin_model_general}, the induced exchange is
$J_{ij}^{(2)}=-J_K^2\Pi_{ij}$; a ferromagnetic nearest-neighbor
interaction corresponds to $J_{ij}^{(2)}>0$.

The scalar-chirality term first appears at third order in $J_K$:
\begin{equation}
    H^{(3)}_{\rm eff}
    =
    \frac{J_K^3}{3}
    \sum_{ijk}
    \mathcal G_{ijk}\,
    {\rm tr}_{\sigma}
    \left[
        (\mathbf S_i\cdot\boldsymbol{\sigma})
        (\mathbf S_j\cdot\boldsymbol{\sigma})
        (\mathbf S_k\cdot\boldsymbol{\sigma})
    \right],
    \label{eq:kondo_cubic_before_trace}
\end{equation}
where
\begin{equation}
    \mathcal G_{ijk}
    =
    T\sum_{\omega_n}
    G_0(i,j;i\omega_n)\,
    G_0(j,k;i\omega_n)\,
    G_0(k,i;i\omega_n).
    \label{eq:electronic_triangle_loop}
\end{equation}
Using
\begin{equation}
    {\rm tr}_{\sigma}
    \left[
        \sigma^a\sigma^b\sigma^c
    \right]
    =
    2i\,\epsilon^{abc},
    \label{eq:pauli_trace_three}
\end{equation}
we obtain
\begin{equation}
    H^{(3)}_{\rm eff}
    =
    \frac{2iJ_K^3}{3}
    \sum_{ijk}
    \mathcal G_{ijk}\,
    \mathbf S_i\cdot
    \left(
        \mathbf S_j\times \mathbf S_k
    \right).
    \label{eq:kondo_cubic_ordered_sum}
\end{equation}
The scalar chirality changes sign under $j\leftrightarrow k$, while
Hermiticity of $H_0$ implies
\begin{equation}
    \mathcal G_{ikj}
    =
    \mathcal G_{ijk}^*
    \label{eq:loop_complex_conjugate}
\end{equation}
for the static loop amplitude.  Therefore only the imaginary,
orientation-odd part of the electronic loop contributes.  Writing the
result as a sum over each oriented elementary triangle gives
\begin{equation}
    H^{(3)}_{\rm eff}
    =
    -
    \sum_{\triangle_{ijk}}
    \gamma^{(K)}_{ijk}\,
    \mathbf S_i\cdot
    \left(
        \mathbf S_j\times \mathbf S_k
    \right),
    \label{eq:kondo_chiral_term}
\end{equation}
with
\begin{equation}
    \gamma^{(K)}_{ijk}
    =
    4J_K^3\,
    {\rm Im}\,\mathcal G_{ijk}.
    \label{eq:kondo_gamma_loop}
\end{equation}
The numerical prefactor in Eq.~\eqref{eq:kondo_gamma_loop} assumes that
$\sum_{\triangle_{ijk}}$ counts each oriented triangle once. 

Eqs.~\eqref{eq:kondo_quadratic_realspace},
\eqref{eq:kondo_chiral_term} give the spin Hamiltonian in:
\begin{equation}
    H_{\rm eff}
    =
    -
    \sum_{ij}
    J^{(2)}_{ij}\,
    \mathbf S_i\cdot \mathbf S_j
    -
    \sum_{\triangle_{ijk}}
    \gamma^{(K)}_{ijk}\,
    \mathbf S_i\cdot
    \left(
        \mathbf S_j\times \mathbf S_k
    \right).
    \label{eq:kondo_spin_hamiltonian}
\end{equation}
Upon restricting to nearest-neighbor exchange and a uniform chiral
coupling, as well as an applied out-of-plane magnetic field which introduces a term $-h\sum_iS_i^z$, the above reduces to Eq.~\eqref{eq:central}.

The origin of ${\rm Im}\,\mathcal G_{ijk}$ is transparent on a single
triangular plaquette.  In a localized or strong-binding limit, the
off-diagonal Green's function is proportional to the hopping matrix
element.  Thus, to leading nonvanishing order in hopping,
\begin{equation}
    \mathcal G_{123}
    =
    C_6\,
    t_{12}t_{23}t_{31}
    +O(t^4),
    \label{eq:plaquette_loop_strong_binding}
\end{equation}
where $C_6$ is a real coefficient determined by the local Green's
function.  For example, for a single atomic level
$g_0(i\omega_n)=(i\omega_n-\epsilon_0)^{-1}$, one has
$C_6\propto T\sum_{\omega_n}(g_0(i\omega_n))^6$, up to a sign fixed by the
hopping convention.  Therefore
\begin{equation}
    {\rm Im}\,\mathcal G_{123}
    =
    C_6\,
    |t_{12}t_{23}t_{31}|\,
    \sin\Phi_{123}.
    \label{eq:imaginary_loop_sin_flux}
\end{equation}
The scalar-chirality interaction vanishes for real hoppings, for which
$\Phi_{123}=0$ or $\pi$, and is maximized when the loop phase is near
$\pm\pi/2$.  Thus the chiral spin interaction is the spin-sector
descendant of a time-reversal-odd orbital loop amplitude.

\subsection{Hubbard Model}

The same structure arises in an itinerant Hubbard model with
spin-independent hopping,
\begin{equation}
    H_{\rm Hub}
    =
    H_0
    +
    U\sum_i n_{i\uparrow}n_{i\downarrow},
    \qquad
    U>0 ,
    \label{eq:hubbard_hamiltonian}
\end{equation}
where $H_0$ is given by Eq.~\eqref{eq:electronic_h0}.  This route is
best viewed as a weak-coupling or Stoner-like effective
magnetic action, showing that even itinerant systems without pre-existing local moments can give rise to the chiral interaction.

Let
\begin{equation}
    \mathbf s_i
    =
    \frac{1}{2}
    c^\dagger_{i\alpha}
    \boldsymbol{\sigma}_{\alpha\beta}
    c_{i\beta}
    \label{eq:electron_spin_physical}
\end{equation}
be the physical electron spin operator.  The single-site operator
identity
\begin{equation}
    n_{i\uparrow}n_{i\downarrow}
    =
    \frac{1}{2}n_i
    -
    \frac{2}{3}\,
    \mathbf s_i^2
    \label{eq:hubbard_spin_identity}
\end{equation}
allows us, at fixed density, to retain only the spin-channel
interaction.  Defining $U_s=2U/3$, the Euclidean action may be written
as
\begin{equation}
    S
    =
    \int_0^\beta d\tau
    \left[
        \sum_{ij,\alpha}
        c^\dagger_{i\alpha}
        \left(
            \partial_\tau\delta_{ij}
            +
            h^0_{ij}
        \right)
        c_{j\alpha}
        -
        U_s
        \sum_i
        \mathbf s_i^2
    \right],
    \label{eq:hubbard_spin_action}
\end{equation}
where $h^0_{ij}$ is the one-particle hopping matrix associated with
$H_0$.

We decouple the spin interaction by a real Hubbard-Stratonovich field
$\mathbf M_i(\tau)$:
\begin{equation}
    \exp\!\left[
        U_s\int d\tau\,\mathbf s_i^2
    \right]
    \propto
    \int d\mathbf M_i\,
    \exp\!\left[
        -\int d\tau
        \left(
            \frac{\mathbf M_i^2}{4U_s}
            -
            \mathbf M_i\cdot\mathbf s_i
        \right)
    \right].
    \label{eq:hs_identity_spin}
\end{equation}
The fermionic action then becomes quadratic,
\begin{equation}
    S[c^\dagger,c,\mathbf M]
    =
    \int_0^\beta d\tau
    \left[
        \sum_{ij}
        c^\dagger_i
        \left(
            G_0^{-1}
            -
            \hat V_M
        \right)_{ij}
        c_j
        +
        \sum_i
        \frac{\mathbf M_i^2}{4U_s}
    \right],
    \label{eq:hubbard_decoupled_action}
\end{equation}
with
\begin{equation}
    (\hat V_M)_{ij}
    =
    \frac{1}{2}
    \delta_{ij}\,
    \mathbf M_i\cdot\boldsymbol{\sigma}.
    \label{eq:hubbard_vertex}
\end{equation}
After integrating out the fermions,
\begin{equation}
    S_{\rm eff}[\mathbf M]
    =
    \int_0^\beta d\tau
    \sum_i
    \frac{\mathbf M_i^2}{4U_s}
    -
    {\rm Tr}\,
    \ln
    \left[
        G_0^{-1}
        -
        \hat V_M
    \right].
    \label{eq:hubbard_effective_action}
\end{equation}
Expanding the logarithm gives the usual magnetic functional \cite{altland2010condensed}.

At quadratic order,
\begin{align}
    \notag &S_{\rm eff}^{(2)}
    = \\ &
    \frac{1}{4}
    \sum_{\mathbf q,\Omega_m}
    \left[
        U_s^{-1}
        -
        \chi_0(\mathbf q,i\Omega_m)
    \right]
    \mathbf M(\mathbf q,i\Omega_m)
    \cdot
    \mathbf M(-\mathbf q,-i\Omega_m),
    \label{eq:hubbard_quadratic_action}
\end{align}
where the bare particle-hole susceptibility is
\begin{equation}
    \chi_0(\mathbf q,i\Omega_m)
    =
    -
    T\sum_{\mathbf k,\omega_n}
    G_0(\mathbf k+\mathbf q,i\omega_n+i\Omega_m)
    G_0(\mathbf k,i\omega_n).
    \label{eq:hubbard_bare_susceptibility}
\end{equation}
The Stoner instability occurs when
\begin{equation}
    U_s\,\chi_0(\mathbf Q,0)=1
    \label{eq:stoner_criterion}
\end{equation}
for the ordering wave vector $\mathbf Q$ that maximizes
$\chi_0(\mathbf q,0)$.  In the ferromagnetic case, this maximum occurs
at $\mathbf Q=0$.

At cubic order, the same electronic loop encountered in the Kondo
problem appears:
\begin{equation}
    S_{\rm eff}^{(3)}
    =
    \frac{1}{3}
    {\rm Tr}
    \left[
        G_0\hat V_M
        G_0\hat V_M
        G_0\hat V_M
    \right].
    \label{eq:hubbard_cubic_trace}
\end{equation}
Using the Pauli trace in Eq.~\eqref{eq:pauli_trace_three} and then
antisymmetrizing over the orientation of the triangle gives
\begin{equation}
    S_{\rm eff}^{(3)}
    =
    -
    \frac{1}{2}
    \sum_{\triangle_{ijk}}
    {\rm Im}\,\mathcal G_{ijk}\,
    \mathbf M_i\cdot
    \left(
        \mathbf M_j\times \mathbf M_k
    \right),
    \label{eq:hubbard_chiral_m_action}
\end{equation}
where $\mathcal G_{ijk}$ is the same loop amplitude defined in
Eq.~\eqref{eq:electronic_triangle_loop}.  Thus the Hubbard model also
generates a scalar-chirality term whenever the electronic loop amplitude
has a nonzero imaginary part.

Below the magnetic transition, if the amplitude of the magnetic order
parameter is approximately fixed, $\mathbf M_i=M_0\mathbf S_i$, the
cubic term becomes
\begin{align}
    \notag H_{\rm Hub}^{(3)}
    &=
    -
    \sum_{\triangle_{ijk}}
    \gamma^{({\rm Hub})}_{ijk}\,
    \mathbf S_i\cdot
    \left(
        \mathbf S_j\times \mathbf S_k
    \right),
    \\ &
    \gamma^{({\rm Hub})}_{ijk}
    =
    \frac{M_0^3}{2}\,
    {\rm Im}\,\mathcal G_{ijk}.
    \label{eq:hubbard_gamma}
\end{align}
The microscopic coefficient is therefore again proportional to
$\sin\Phi_{ijk}$ in the plaquette limit.  The Hubbard example makes
clear that the same chiral interaction can arise from an itinerant
Stoner instability in a band structure with time-reversal-odd orbital
phases or Berry curvature.

\section{Gradient expansion of the Heisenberg term on the triangular lattice}
\label{app:heisenberg}

In this appendix we derive the continuum expansion of the nearest-neighbor Heisenberg interaction
\begin{equation}
H_{\rm Heis}
=
-J\sum_{\langle ij\rangle}\mathbf n_i\cdot \mathbf n_j
\label{eq:App_HHeis_start}
\end{equation}
for a slowly varying unit vector field \(\mathbf n(\mathbf r)\).

We take the triangular lattice primitive vectors
\begin{equation}
\mathbf a_1=(1,0),
\qquad
\mathbf a_2=\left(-\frac12,\frac{\sqrt3}{2}\right),
\label{eq:App_tri_vectors}
\end{equation}
and nearest-neighbor displacements
\begin{equation}
\boldsymbol\delta\in
\{\pm \mathbf a_1,\pm \mathbf a_2,\pm(\mathbf a_1+\mathbf a_2)\}.
\label{eq:App_deltas}
\end{equation}
The bond sum can be written as
\begin{equation}
H_{\rm Heis}
=
-\frac{J}{2}\sum_{{\mathbf r}, {\boldsymbol\delta}}
\mathbf n(\mathbf r)\cdot \mathbf n(\mathbf r+\boldsymbol\delta),
\label{eq:App_bondsum}
\end{equation}
where the factor \(1/2\) avoids double counting.

For slowly varying fields,
\begin{equation}
\begin{aligned}
\mathbf n(\mathbf r+\boldsymbol\delta)
&=
\mathbf n
+(\boldsymbol\delta\cdot\nabla)\mathbf n
+\frac12(\boldsymbol\delta\cdot\nabla)^2\mathbf n
\\
&\quad
+\frac16(\boldsymbol\delta\cdot\nabla)^3\mathbf n
+\frac1{24}(\boldsymbol\delta\cdot\nabla)^4\mathbf n
\\
&\quad
+\frac1{720}(\boldsymbol\delta\cdot\nabla)^6\mathbf n
+\cdots
\end{aligned}
\label{eq:App_Taylor_Heis}
\end{equation}
Odd-order terms vanish after summing over \(\pm\boldsymbol\delta\).  Using \(\mathbf n^2=1\), one has the standard identities
\begin{equation}
\mathbf n\cdot \partial_\mu \mathbf n =0,
\qquad
\mathbf n\cdot \partial_\mu\partial_\nu \mathbf n
=
-(\partial_\mu \mathbf n)\cdot (\partial_\nu \mathbf n).
\label{eq:App_unitconstraint}
\end{equation}
Substituting Eq.~\eqref{eq:App_Taylor_Heis} into Eq.~\eqref{eq:App_bondsum} and keeping terms through fourth order in gradients gives
\begin{equation}
\begin{aligned}
H_{\rm Heis}
={}&
E_0
+\frac{J}{4}\sum_{\mathbf r}\sum_{\boldsymbol\delta}
\delta_\mu\delta_\nu\,
(\partial_\mu \mathbf n)\cdot(\partial_\nu \mathbf n)
\\
&\;
-\frac{J}{48}\sum_{\mathbf r}\sum_{\boldsymbol\delta}
\delta_\mu\delta_\nu\delta_\rho\delta_\sigma\,
(\partial_\mu\partial_\nu \mathbf n)\cdot
(\partial_\rho\partial_\sigma \mathbf n)
+\cdots
\end{aligned}
\label{eq:App_Heis_expand1}
\end{equation}
where \(E_0\) is the ferromagnetic constant.

The second-order tensor obtained from the triangular-lattice nearest-neighbor shell is isotropic:
\begin{equation}
\sum_{\boldsymbol\delta}\delta_\mu\delta_\nu
=
3\,\delta_{\mu\nu}.
\label{eq:App_secondmoment}
\end{equation}
Using \(\sum_{\mathbf r}\to A_{\rm uc}^{-1}\int d^2r\), this gives
\begin{equation}
H_{\rm Heis}^{(2)}
=
\frac{3J}{4A_{\rm uc}}
\int d^2r\, (\partial_\mu \mathbf n)^2.
\label{eq:App_Heis_2nd}
\end{equation}

The fourth-order tensor also reduces to the isotropic combination appropriate to the triangular lattice.  After carrying out the neighbor sum and integrating by parts, one obtains
\begin{equation}
H_{\rm Heis}^{(4)}
=
-\frac{3J}{64A_{\rm uc}}
\int d^2r\, (\nabla^2 \mathbf n)^2.
\label{eq:App_Heis_4th}
\end{equation}
Combining the two pieces,
\begin{equation}
\begin{aligned}
H_{\rm Heis}
={}&
E_0
+\frac{3J}{4A_{\rm uc}}
\int d^2r\, (\partial_\mu \mathbf n)^2
\\
&\;
-\frac{3J}{64A_{\rm uc}}
\int d^2r\, (\nabla^2 \mathbf n)^2
+\mathcal O(\partial^6).
\end{aligned}
\label{eq:App_Heis_final}
\end{equation}

To expose the scaling with the skyrmion size, let
$\mathbf{x}=\mathbf{r}/\lambda$ and
$\mathbf{n}(\mathbf{r})=\mathbf{n}_0(\mathbf{x})$. Then
\begin{equation}
\begin{aligned}
\int d^2r\,(\partial_\mu \mathbf{n})^2
&=
\int d^2x\,(\partial_{x_\mu}\mathbf{n}_0)^2,
\\
\int d^2r\,(\nabla^2 \mathbf{n})^2
&=
\frac{1}{\lambda^2}
\int d^2x\,(\nabla_{\mathbf{x}}^2\mathbf{n}_0)^2.
\end{aligned}
\label{eq:App_scaling_Heis}
\end{equation}
The corresponding magnetic-unit-cell area scales as
$A_\lambda=\lambda^2 A_0$. Therefore, after dividing the
total energy by $A_\lambda$, the two Heisenberg contributions
generate terms proportional to $1/\lambda^2$ and
$1/\lambda^4$, respectively, in the expansion of the energy
density.

For completeness, the sixth-order part has the anisotropic triangular-lattice structure
\begin{equation}
\begin{aligned}
H_{\rm Heis}^{(6)}
={}&
\frac{J}{7680A_{\rm uc}}
\int d^2r\,\Big[
11(\partial_x^3\mathbf n)^2
+9(\partial_y^3\mathbf n)^2
\\
&\quad
+15(\partial_x^2\partial_y\mathbf n)^2
+45(\partial_x\partial_y^2\mathbf n)^2
\Big].
\end{aligned}
\label{eq:App_Heis_6th}
\end{equation}
The sixth-order terms therefore contribute at order
$1/\lambda^4$ to the total energy, or equivalently at order
$1/\lambda^6$ to the energy density.

\section{Gradient expansion of the chiral term on the triangular lattice}
\label{app:chiral}

We now derive the continuum limit of the scalar-chirality interaction
\begin{equation}
H_{\chi}
=
-\gamma\sum_{\triangle ijk}
\mathbf n_i\cdot(\mathbf n_j\times \mathbf n_k).
\label{eq:App_Hchi_start}
\end{equation}
Each elementary triangle contributes the oriented solid angle formed by the three neighboring spins.

To organize the expansion, it is convenient to rewrite the plaquette sum as a site sum over the six triangles surrounding each lattice site:
\begin{equation}
H_{\chi}
=
-\frac{\gamma}{3}\sum_{\mathbf r}
\sum_{(\boldsymbol\delta_1,\boldsymbol\delta_2)}
\mathbf n(\mathbf r)\cdot
\left[
\mathbf n(\mathbf r+\boldsymbol\delta_1)
\times
\mathbf n(\mathbf r+\boldsymbol\delta_2)
\right],
\label{eq:App_Hchi_sitesum}
\end{equation}
where the pairs \((\boldsymbol\delta_1,\boldsymbol\delta_2)\) run over the six oriented edge pairs of the elementary triangles, and the factor \(1/3\) compensates the triple counting of each plaquette.

We now expand both shifted spins about \(\mathbf r\):
\begin{equation}
\begin{aligned}
\mathbf n(\mathbf r+\boldsymbol\delta)
={}&
\mathbf n
+(\boldsymbol\delta\cdot\nabla)\mathbf n
+\frac12(\boldsymbol\delta\cdot\nabla)^2\mathbf n
\\
&\;
+\frac16(\boldsymbol\delta\cdot\nabla)^3\mathbf n
+\frac1{24}(\boldsymbol\delta\cdot\nabla)^4\mathbf n
+\cdots
\end{aligned}
\label{eq:App_Taylor_chi}
\end{equation}
Substituting into Eq.~\eqref{eq:App_Hchi_sitesum}, the zeroth- and first-order terms vanish identically, and the first nonzero contribution is second order in gradients:
\begin{equation}
H_{\chi}^{(2)}
=
-\frac{\gamma}{3A_{\rm uc}}
\int d^2r
\sum_{(\boldsymbol\delta_1,\boldsymbol\delta_2)}
\mathbf n\cdot
\left[
(\boldsymbol\delta_1\cdot\nabla)\mathbf n
\times
(\boldsymbol\delta_2\cdot\nabla)\mathbf n
\right].
\label{eq:App_Hchi_second_raw}
\end{equation}
The lattice sum can be performed explicitly for the triangular geometry and gives
\begin{equation}
H_{\chi}^{(2)}
=
-2\gamma
\int d^2r\,
\mathbf n\cdot
(\partial_x\mathbf n\times \partial_y\mathbf n).
\label{eq:App_Hchi_second}
\end{equation}
Using the definition of the skyrmion number,
\begin{equation}
N_{\rm sk}
=
\frac{1}{4\pi}\int d^2r\,
\mathbf n\cdot
(\partial_x\mathbf n\times \partial_y\mathbf n),
\label{eq:App_Qdef}
\end{equation}
this becomes
\begin{equation}
H_{\chi}^{(2)}=-8\pi\gamma\, N_{\rm sk}.
\label{eq:App_Hchi_Q}
\end{equation}
This is the microscopic origin of the topological chemical potential discussed in the main text.

To obtain the scale-selecting corrections, we must continue the expansion to fourth order in spatial derivatives.  Collecting the terms of total derivative order four gives
\begin{equation}
\begin{aligned}
H_{\chi}^{(4)}
={}&
-\frac{\gamma}{3A_{\rm uc}}
\int d^2r
\sum_{(\boldsymbol\delta_1,\boldsymbol\delta_2)}
\mathbf n\cdot\Big[
\frac14
(\boldsymbol\delta_1\!\cdot\!\nabla)^2\mathbf n
\times
(\boldsymbol\delta_2\!\cdot\!\nabla)^2\mathbf n
\\
&\quad
+\frac16
(\boldsymbol\delta_1\!\cdot\!\nabla)\mathbf n
\times
(\boldsymbol\delta_2\!\cdot\!\nabla)^3\mathbf n
\\
&\quad
+\frac16
(\boldsymbol\delta_1\!\cdot\!\nabla)^3\mathbf n
\times
(\boldsymbol\delta_2\!\cdot\!\nabla)\mathbf n
\Big].
\end{aligned}
\label{eq:App_Hchi_fourth_raw}
\end{equation}
After explicitly summing over the six oriented triangles and simplifying by antisymmetry of the vector product, one obtains the quartic correction in the form
\begin{equation}
H_{\chi}^{(4)}
=
-\frac{\gamma\sqrt3}{16A_{\rm uc}}
\int d^2r\, I_4[\mathbf n],
\label{eq:App_Hchi_fourth}
\end{equation}
where
\begin{equation}
\begin{aligned}
I_4[\mathbf n]
={}&
\mathbf n\!\cdot\!\Big[
\partial_x\mathbf n\times \partial_y^3\mathbf n
+\partial_x\mathbf n\times \partial_x^2\partial_y\mathbf n
\\
&\quad
+\partial_x\partial_y\mathbf n\times \partial_y^2\mathbf n
+\partial_x\partial_y^2\mathbf n\times \partial_y\mathbf n
\\
&\quad
+\partial_x^2\mathbf n\times \partial_x\partial_y\mathbf n
+\partial_x^3\mathbf n\times \partial_y\mathbf n
\Big]
\\
&\;
-\bigl(\text{same terms with the two factors exchanged}\bigr).
\end{aligned}
\label{eq:App_I4}
\end{equation}
This is the explicit four-derivative chiral invariant reported in the
main text [Eq.~\eqref{eq:Chiral_grad_main}]. As a local energy-density
contribution, $I_4[\mathbf n]$ scales as $1/\lambda^4$, whereas its
spatial integral scales as $1/\lambda^2$. It therefore contributes
directly to the coefficient $b$ in the energy-density expansion.

Thus, up to fourth order in gradients,
\begin{equation}
H_{\chi}
=
-2\gamma\int d^2r\,
\mathbf n\cdot(\partial_x\mathbf n\times \partial_y\mathbf n)
-\frac{\gamma\sqrt3}{16A_{\rm uc}}\int d^2r\, I_4[\mathbf n]
+\mathcal O(\partial^6).
\label{eq:App_Hchi_final}
\end{equation}
For a texture $\mathbf{n}(\mathbf{r})=
\mathbf{n}_0(\mathbf{x})$, with $\mathbf{x}=\mathbf{r}/\lambda$,
the two contributions to the total energy scale as
\begin{equation}
\begin{aligned}
\int d^2r\,
\mathbf{n}\cdot
\left(\partial_x\mathbf{n}\times\partial_y\mathbf{n}\right)
&=
\int d^2x\,
\mathbf{n}_0\cdot
\left(
\partial_{x_1}\mathbf{n}_0
\times
\partial_{x_2}\mathbf{n}_0
\right),
\\
\int d^2r\,I_4[\mathbf{n}]
&=
\frac{1}{\lambda^2}
\int d^2x\,I_4[\mathbf{n}_0].
\end{aligned}
\label{eq:App_scaling_chi}
\end{equation}
The first integral is scale invariant, as expected for the
topological charge, whereas the quartic correction scales as
$1/\lambda^2$. Since the magnetic-unit-cell area scales as
$A_\lambda=\lambda^2 A_0$, the corresponding contributions
to the energy density scale as $1/\lambda^2$ and
$1/\lambda^4$, respectively.

\section{Numerical Calculations}

The behavior near the phase boundary and the global phase diagram shown in Fig.~\ref{fig:shortwavelength_pd}~(a) were determined through the application of several complementary methods. Unbiased optimization and annealing efforts in the long-wavelength portion of the phase diagram ($\gamma/J < 1.3$) failed to yield any spin configuration with energy lower than the variational ansatz discussed in Sec.~\ref{sec:dlb_skx_ansatz}, namely the double skyrmion spin texture prescribed by Eq.~\eqref{eq:tripleQansatz} subjected to numerical gradient optimization. Consequently, we relied on this gradient-optimized ansatz for all analysis close to $\gamma_m/J$, in the first case by fitting parameters to the energy density Eq.~\eqref{eq:exp} and using the expression Eq.~\eqref{eq:hsat} [Appendix~\ref{sec:app_coeff_fitting}]. The phase boundary that results from this approach is shown in Fig.~\ref{fig:shortwavelength_pd} (a) as a solid magenta line. To determine the evolution of the phase diagram as $(\gamma-\gamma_m)/J$ grows and field is applied, we continued to rely on this ansatz in the region 
$\gamma_m/J < \gamma/J < 1.4$ by again optimizing the double skyrmion ansatz numerically, in this case also for a range of applied field values. Appropriate interpolations were made between simulations performed on a large number of system sizes to account for the quasi-continuous evolution of the skyrmion lengthscale $\lambda$ [Appendix~\ref{sec:app_num_lw_ansatz}]. The results of this study are shown in Fig.~\ref{fig:unbiased_near_gammam}, and the phase boundary is shown in the global phase diagram of Fig.~\ref{fig:shortwavelength_pd} (a) as a red line. We note that the long-wavelength theory agrees with these numerical results between 
$\gamma/J=\gamma_m/J$ and $\gamma/J\approx1.2$. To determine the phase diagram away from the continuum limit ($\gamma/J \geq 1.3$), we employed an unbiased simulated annealing and gradient optimization approach on variable-size supercells [Appendix~\ref{sec:app_num_pd}].

~\subsection{Coefficients of Energy Expansion\label{sec:app_coeff_fitting}}

\begin{figure}[htb]
    \centering
    \includegraphics[width=1.0\columnwidth]{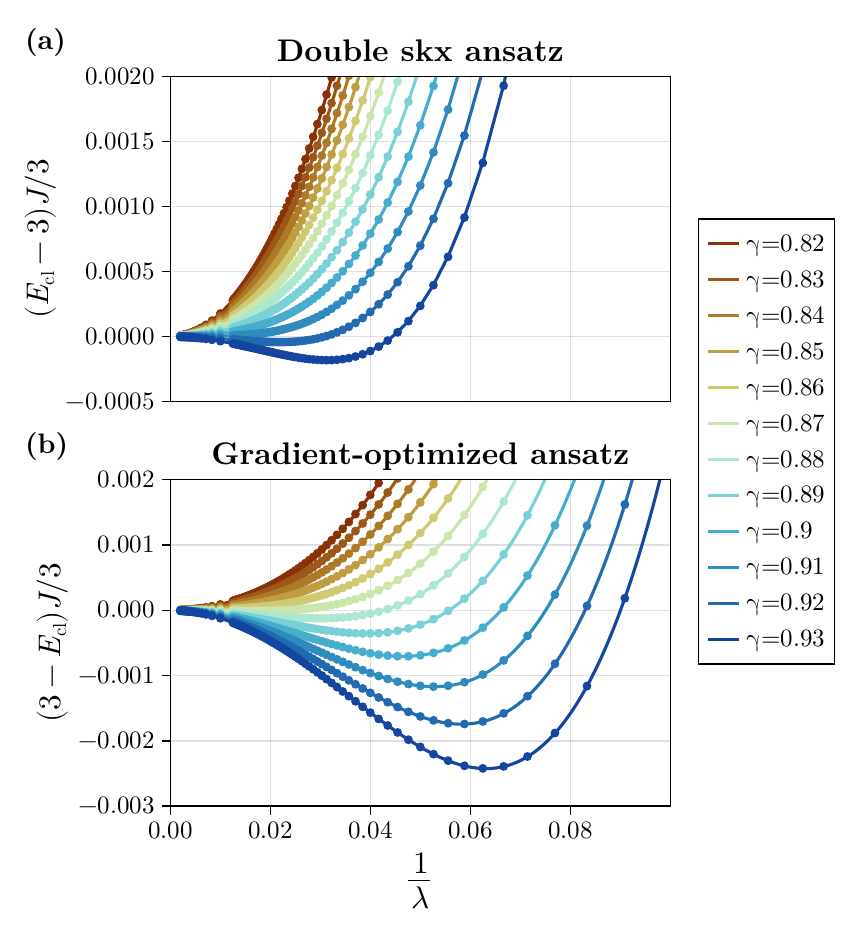}
    \caption{
    Classical energies (dots) and 6th-order polynomial fit to energies (lines) as evaluated for (a) the ansatz of Eq.~\eqref{eq:tripleQansatz} and (b) the gradient-optimized ansatz. Colors represent different $\gamma$ values, which were selected in proximity to the critical point $\gamma_m=\sqrt{3}/2$. The energy of the lattice Hamiltonian Eq.~\eqref{eq:central} was offset by $3J$ and divided by $3$ to put the energy scale of the microscopic Hamiltonian [Eq.~\eqref{eq:central}] in correspondence with the continuum model Eq.~\eqref{eq:Hcont_mu}.}
    \label{fig:energy_polynomial_fitting}
\end{figure}

As $\gamma - \gamma_m$ becomes small, the characteristic lengthscale of the skyrmion solutions becomes arbitrarily large, making unbiased numerical simulation difficult. To examine the phase diagram in this region, we rely on the numerically relaxed ansatz described in Sec.~\ref{sec:dlb_skx_ansatz}. To fit the coefficients of Eq.~\eqref{eq:exp}, this ansatz was initialized with values of $\lambda$ on a lattice of linear dimension $L=\lambda$ for select integer values between $L=10$ and $L=500$ and $\gamma/J$ values in increments of $0.01$ between $0.82$ and $0.93$. The ansatz itself, Eq.~\eqref{eq:tripleQansatz}, was examined together with a spin configuration arrived at by applying numerical gradient optimization to the ansatz. For each case the results for all $\lambda$ and $\gamma$ values were used to fit the parameters of the expression:
 \begin{equation}
e = \left(a_{\rm h} + \gamma a_{\chi}\right)\frac{1}{\lambda^2} + \left(b_{\rm h} + \gamma b_{\chi}\right)\frac{1}{\lambda^4} + \left(c_{\rm h} + \gamma c_{\chi}\right)\frac{1}{\lambda^6}.
\label{eq:energydensitywithgamma}
\end{equation}
Here we have simply expanded the coefficients into Heisenberg contribution (subscript $h$) and chiral contribution (subscript ${\chi}$), which are always related as $a_h +\gamma a_{\chi}$, etc.  The results of these fits are illustrated in Fig.~\ref{fig:energy_polynomial_fitting}. The resulting coefficients are summarized in Table \ref{tab:energy_fits}.

\begin{table}
\begin{tabular}{|c|c|c|c|c|c|c|}
\hline 

& $a_{\rm h}$ & $a_{\chi}$ & $b_{\rm h}$ & $b_{\chi}$ & $c_{\rm h}$ & $c_{\chi}$ \tabularnewline
\hline 
\hline 
Ansatz & 17.56 & -19.28  & -95.62 & 311.74   & -1071.17  & -1744.65 \tabularnewline
\hline 
Opt. Ansatz& 16.73 & -19.32  & -62.24 & 241.05 &  -638.44 & -590.68 \tabularnewline
\hline

\end{tabular}
\caption{Least-squares fit of parameters of Eq.~\eqref{eq:energydensitywithgamma} for the ansatz (Ansatz) of Eq.~\eqref{eq:tripleQansatz} and gradient-optimized ansatz (Opt. Ansatz).}
\label{tab:energy_fits}
\end{table}

The lower bounds on quadratic coefficients are known analytically from the nonlinear sigma model and are given by $a_{\rm h}=6\pi/3\approx16.76$ and $a_{\chi}=-32\sqrt{3}\pi/9\approx-19.35$. Both these limits are saturated by the gradient-optimized ansatz but not the unoptimized ansatz. (In fact, the optimized ansatz results are very slightly below  this limit, an effect we attributed to numerical approximation and fitting uncertainty.) At $\gamma=\gamma_m$, $b$  is strongly positive ($b\approx146.52$), consistent with a continuous transition and a skyrmion density that grows smoothly with $\gamma$.

\subsection{Numerical calculations with ansatz near phase boundary\label{sec:app_num_lw_ansatz}}

\begin{figure}[htb]
    \centering
    \includegraphics[width=1.0\columnwidth]{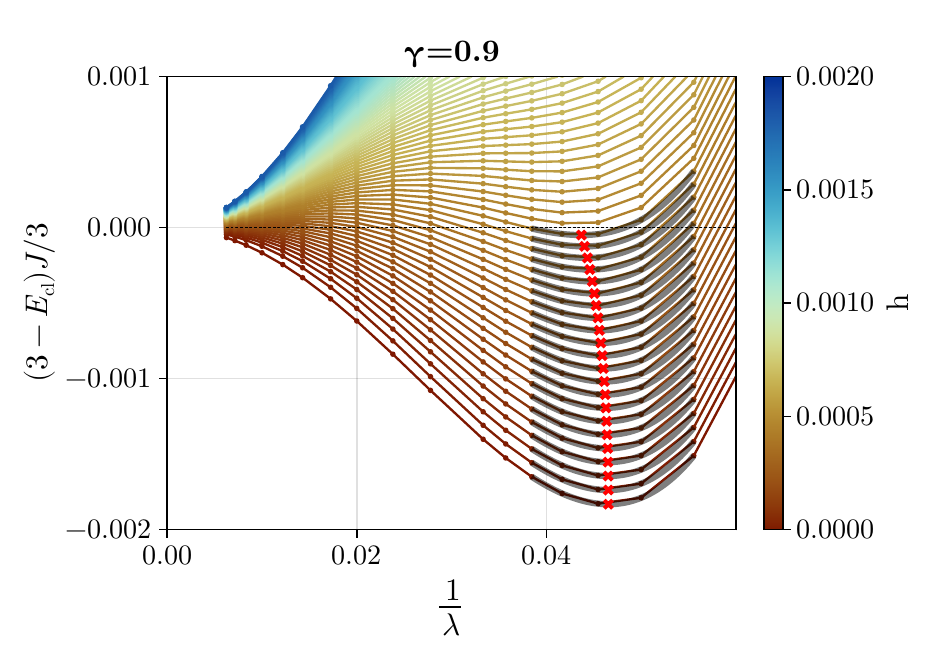}
    \caption{
    Energy of the gradient-optimized double-skrymion ansatz as a function of inverse lengthscale $1/\lambda$. Colors represent different applied field values $h$. For the set of fields where the lowest energies are less than the ferromagnetic energy (dashed black line), a third-order spline interpolation was performed about the anticipated minimum. The minimum of this interpolant was determined and taken as the optimal length scale for each $h$ (red crosses). The figure shows only the results for $\gamma=0.9$. The combined results for a series of $\gamma/J$ values are shown in Fig.~\ref{fig:unbiased_near_gammam} (a).
    }
    \label{fig:longwavelength_interpolation}
\end{figure}

The approach to fitting the energy density parameters just outlined was also used  to investigate the evolution of  the ordering wave vector and magnetization in the long-wavelength portion of the phase diagram. For a range of $\gamma$ values between 0.9 and 1.4, the double skyrmion ansatz Eq.~\eqref{eq:tripleQansatz} with lengthscale $\lambda$ was initialized on a supercell of linear dimension $L=\lambda$ for a range of $\lambda$s and applied fields $h$. Each spin configuration was then subject to numerical optimization via the conjugate-gradient algorithm implemented in the Optim.jl package \cite{mogensen2018optim} using the exact gradient of Eq.~\eqref{eq:central}. The energy and magnetization were recorded for the minimized results. For each $\gamma$ and $h$, a third-order spline was fit to the energy values in proximity of the expected global minimum to determine the optimal skyrmion lengthscale $\lambda^*$. The corresponding magnetization was determined by interpolating between the magnetization values corresponding to the same $\lambda$ values and determining the value of the interpolant at $\lambda^*$. An illustration of this procedure is shown in Fig.~\ref{fig:longwavelength_interpolation}.

\subsection{Calculation of short-wavelength phase diagram\label{sec:app_num_pd}}

To capture the main features of the phase diagram away from $\gamma_m$, unbiased numerical simulations were performed over a grid of $\gamma_j$ values between 1.3 and 5 $J$ in increments of 0.1, and $h_k$ values between 0 and 2.5 $J$, also in increments of 0.1. For each pair of values, a Hamiltonian $H_{jk}$ was established with $J=1$, $\gamma=\gamma_j J$ and $h=h_k J$. Each $H_{ij}$ was then associated with supercell of linear dimension $L$, where $L$ was selected randomly from values between 3 and 50. The spin configuration on each supercell, ${\bm S}_{jk}$ was initialized with random (infinite temperature) initial conditions and subjected to a brief annealing procedure: using the stochastic Landau-Lifshitz-Gilbert dynamics \cite{brown1963thermal} with an empirical damping parameter of $\mu=0.1$, the dynamics were integrated numerically using the Stochastic-Heun method and a time step of $\Delta t=0.01 J^{-1}$ for 5000 steps. A geometric cooling schedule was applied such that the temperature at time step $n$ was set to $T_n=0.9989^n$. After annealing, each spin configuration was further optimized directly via gradient descent using the conjugate gradient algorithm provided by the Optim.jl package \cite{mogensen2018optim}. After the generation of initial candidates, a series of refinement iterations were performed as follows. For each point on the phase diagram $\left(\gamma_i,h_j\right)$, the same procedure used to determine the initial candidates was repeated, and if a lower-energy candidate was found, it was retained. Additionally, for each $H_{jk}$, the spin configurations of the neighboring points of the phase diagram, ${\bm S}_{j\pm1, k\pm1}$, were taken as initial conditions for $H_{jk}$ and subjected to gradient optimization. If any of the resulting configurations had lower energy than the current best candidate for $H_{jk}$, it was retained as the new candidate. Iterations over all points of the phase diagram were repeated until convergence, which was taken as 100 consecutive passes without the discovery of any lower energy spin configurations, achieved after approximately 506 passes.  We note this is the approach described in \cite{zhang2023cp2} with the added complication that the dimensions of each supercell were subject to random sampling. The phase boundaries (orange lines) and skyrmion densities (color gradient) determined from the final set of candidates are shown in Fig.~\ref{fig:shortwavelength_pd}. Where this unbiased analysis overlaps with the analysis performed with the ansatz ($1.3 < \gamma/J < 1.5$), agreement is very good.

\section{Explicit Parametrization of the Tetra-Skyrmion State \label{sec:app_tretra-SkX}}

The static spin structure factor of the tetra-skyrmion state, shown in Fig.~\ref{fig:tetraskx_tetrahedral}~(d), exhibits dominant weight at six symmetry-related wave vectors located midway between the $\Gamma$ point and the corners (K points) of the hexagonal Brillouin zone, together with subdominant contributions at the M points corresponding to the midpoints of the zone edges. Because of this simple structure, the phase admits an analytical expression.
The spin configuration can be parametrized in real space by a $24$-site magnetic unit cell defined by the lattice vectors [see Figs.~\ref{fig:tetraskx_tetrahedral} (b) and (c)]:
\begin{equation}
\mathbf T_1 = 6 \mathbf a_1, \qquad 
\mathbf T_2 = 2 \mathbf a_1 + 4 \mathbf a_2.
\end{equation}
Labeling the sites within the unit cell by
\begin{equation}
\mathbf r_{mn} = m \mathbf a_1 + n \mathbf a_2,
\qquad m=0,\dots,5,\quad n=0,\dots,3,
\end{equation}
the spin configuration is given by
\begin{equation}
\mathbf n(\mathbf r_{mn}) =
\Bigl(
\sqrt{1 - z_{mn}^2}\,\cos\phi_{mn},\;
\sqrt{1 - z_{mn}^2}\,\sin\phi_{mn},\;
z_{mn}
\Bigr),
\label{eq:realspace_spin}
\end{equation}
with longitudinal components 
\begin{equation}
z_{nm}
=\frac{1}{3}
\left(
\begin{array}{rrrrrr}
-1 & -1 &  1 & -1 & -1 &  1 \\
-1 & -3 & -1 &  3 &  1 &  3 \\
 1 & -1 & -1 &  1 & -1 & -1 \\
-1 &  3 &  1 &  3 & -1 & -3
\end{array}
\right),
\label{eq:Znm_tetraSkX}
\end{equation}
and azimuthal angles
\begin{equation}
\phi_{nm} =
\begin{pmatrix}
\frac{4\pi}{3} & \frac{2\pi}{3} & \frac{\pi}{3} & \frac{2\pi}{3} & \frac{4\pi}{3} & \frac{5\pi}{3} \\
0 & \ast & 0 & \ast & \pi & \ast \\
\frac{\pi}{3} & \frac{2\pi}{3} & \frac{4\pi}{3} & \frac{5\pi}{3} & \frac{4\pi}{3} & \frac{2\pi}{3} \\
0 & \ast & \pi & \ast & 0 & \ast
\end{pmatrix}.
\label{eq:phinm_tetraSkX}
\end{equation}
Here \(\ast\) denotes sites with \(z_{mn}=\pm1\), for which the in-plane component vanishes and the azimuthal angle is undefined.

The spin configuration in Eq.~\eqref{eq:realspace_spin} also admits a compact representation in momentum space dominated by six symmetry-related wave vectors
\begin{equation}
\pm \mathbf Q_1,\quad \pm \mathbf Q_2,\quad \pm \mathbf Q_3,
\end{equation}
with
\begin{equation}
\mathbf Q_1=\frac{\mathbf b_1+\mathbf b_2}{6},\;\;\;
\mathbf Q_2=\frac{2\mathbf b_1-\mathbf b_2}{6},\;\;\;
\mathbf Q_3=\frac{-\mathbf b_1+2\mathbf b_2}{6},
\end{equation}
which satisfy $\mathbf Q_1+\mathbf Q_2+\mathbf Q_3=0$ and are located midway between the $\Gamma$ point and the corners of the hexagonal Brillouin zone. The subleading Fourier components appear at the three $M$ points given in Eq.~\eqref{eq:Mpoints}. A compact form of the Fourier expansion is then obtained by introducing
\begin{equation}
n_+(\mathbf r)\equiv n_x(\mathbf r)+i n_y(\mathbf r).
\end{equation}
The in-plane components can then be written as
\begin{eqnarray}
n_+(\mathbf r) &=&
\sum_{\nu=1}^3
\left[
A_\nu e^{-i\mathbf Q_\nu\cdot\mathbf r}
+
B_\nu e^{i\mathbf Q_\nu\cdot\mathbf r}
\right]
+
\sum_{\nu=1}^3 C_\nu \cos(\mathbf M_\nu\cdot\mathbf r),
\nonumber \\
\end{eqnarray}
where
\[
A_\nu=\frac{2\sqrt2}{9}e^{i\alpha_\nu},\qquad
(\alpha_1,\alpha_2,\alpha_3)=
\left(-\frac{\pi}{3},\frac{2\pi}{3},-\frac{2\pi}{3}\right),
\]
and
\begin{eqnarray}
B_1 &=& e^{i\pi/3}A_2,\qquad
B_2=A_3,\qquad
B_3=e^{i\pi/3}A_1,
\nonumber \\
C_1 &=& \frac12 A_1,\qquad
C_2=\frac{e^{i\pi/3}}{2}A_2,\qquad
C_3=\frac12 A_3.
\end{eqnarray}
The longitudinal component is
\begin{eqnarray}
n_z(\mathbf r) &=&
\frac{2}{9}\Big[
\cos(2\pi\mathbf Q_1\cdot\mathbf r)
-\cos(2\pi\mathbf Q_2\cdot\mathbf r)
-\cos(2\pi\mathbf Q_3\cdot\mathbf r)
\Big]
\nonumber \\
&-& \frac{2}{3\sqrt3}
\sum_{\nu=1}^3 \sin(2 \pi\mathbf Q_\nu\cdot\mathbf r)
\nonumber\\
&-& \frac{1}{9}\Big[
\cos(2\pi{\mathbf M}_1\cdot\mathbf r)
+\cos(2\pi{\mathbf M}_2\cdot\mathbf r)
-\cos(2\pi{\mathbf M}_3\cdot\mathbf r)
\Big].
\nonumber \\
\end{eqnarray}
The full spin texture is therefore
\begin{equation}
\mathbf n(\mathbf r)=
\bigl(\Re n_+(\mathbf r),\,\Im n_+(\mathbf r),\,n_z(\mathbf r)\bigr).
\end{equation}
for the real space configuration defined by Eqs.~\eqref{eq:realspace_spin}, ~\eqref{eq:Znm_tetraSkX} and ~\eqref{eq:phinm_tetraSkX}.
This compact Fourier representation highlights the multi-$\mathbf Q$ character of the tetra-skyrmion crystal, with dominant weight carried by the first harmonics at $\mathbf Q_\nu$ and weaker contributions at the $M$ points.

\section{Calculation of the Dynamical Spin Structure Factor
\label{app:DSSF}}

The dynamical spin structure factor (DSSF) is defined as
\begin{multline}
S_{\rm cl}^{\alpha\beta}(\mathbf q,\omega)
= \, \\
\frac{1}{2\pi N_s} \sum_{i,j} \int_{-\infty}^{\infty}\! dte^{i\omega t-i\mathbf q\cdot\mathbf r_{ij}}
\left\langle S_i^\alpha(t)S_j^\beta(0)\right\rangle_{\rm cl},
\label{eq:DSSF_classical}
\end{multline}
where $N_s$ is the number of lattice sites and the brackets
denote an average over thermal initial conditions. For each point of the phase diagram illustrated in Fig.~\ref{fig:dynamics}, the DSSF was calculated as follows. The lowest energy $T=0$ spin configuration found through the procedure outlined in Appendix \ref{sec:app_num_pd} was extended: starting from a linear dimension $L$, the supercell was repeated periodically $n$ times where $n$ is the smallest integer such that $nL>100$. This ensured good momentum-space resolution. The enlarged spin configuration was thermalized at $T=0.03J$ using Langevin dynamics with time step $\Delta t=0.01J^{-1}$. We generated $1000$ independently thermalized
configurations and used each as an initial condition for a dissipationless Landau--Lifshitz trajectory governed by
\begin{equation}
\frac{d\mathbf S_i}{dt}
=
\mathbf S_i\times\mathbf h_i^{\rm eff},
\qquad
\mathbf h_i^{\rm eff}
=
-\frac{\partial\mathcal H}{\partial\mathbf S_i}.
\label{eq:LL_DSSF}
\end{equation}
These trajectories were calculated with a time step of $\Delta t=\pi/315 \approx 0.00997 J^{-1}$, a value chosen for numerical stability. 800 temporal snapshots were recorded for each trajectory. These were collected at intervals of nine time steps, yielding in a sampling rate of $\Delta t_{\rm samp} = \pi/35 \approx 0.089760 J^{-1}$. This subsampling was chosen to yield 401 non-negative resolved energies between zero and $E_{\max}=(2\pi/\Delta t_{\rm samp})/2=35 J$. Each trajectory was Fourier transformed, both on the lattice and in time, with a Fast Fourier Transform. The Fourier transformed spin-spin correlations were finally calculated from the Fourier transformed trajectories via the convolution theorem. Averaging the 1000 samples generated in this manner yielded a numerical approximation of $S_{\rm cl}^{\alpha\beta}(\mathbf q, \omega)$. 

In the low-temperature harmonic regime, the linearized
Landau--Lifshitz equations yield the same normal-mode frequencies as
linear spin-wave theory. The classical and quantum spectral weights,
however, differ because of their respective thermal occupation
factors. We convert the classical DSSF into the corresponding quantum
spin-wave response according to
\begin{equation}
S_{\rm qm}^{\alpha\beta}(\mathbf q,\omega)
=
\frac{\beta\omega}{1-e^{-\beta\omega}}\,
S_{\rm cl}^{\alpha\beta}(\mathbf q,\omega),
\label{eq:DSSF_quantum}
\end{equation}
where $\beta=1/T$ and $\hbar=k_{\rm B}=1$. This
quantum--classical correspondence factor restores quantum detailed
balance and converts the classical thermal spectral weight into the
corresponding quantum one-magnon spectral weight \cite{schofield1960space, zhang2019dynamical, dahlbom2024quantum}.

The quantity shown in the top panel of each labeled pair in Fig.~\ref{fig:dynamics} is the spin-summed response
\begin{equation}
S_{\rm qm}(\mathbf q,\omega)
=
\sum_{\alpha=x,y,z}
S_{\rm qm}^{\alpha\alpha}(\mathbf q,\omega).
\label{eq:DSSF_trace}
\end{equation}
This corresponds roughly to the information collected in an inelastic neutron scattering experiment; greater fidelity to experiment would demand consideration of magnetic form factors and neutron polarization. The bottom figure of each labeled pair shows the instantaneous (or equal-time) structure factor,
\begin{equation}
S_{\rm qm}\left(\mathbf q\right) = \int d\omega S_{\rm qm}(\mathbf q, \omega),
\end{equation}
giving an indication of what would be observed in a neutron diffraction experiment.

These calculations were performed using a version of the Sunny.jl package \cite{Dahlbom2025}, modified to include the three-spin chiral potential term $\mathbf{S}_i\cdot\left(\mathbf{S}_j\times\mathbf{S}_k\right)$.
\end{document}